\documentclass[showpacs,showkeys,preprintnumbers]{revtex4}

\usepackage[dvips]{graphicx}        
\usepackage{amsmath,amsthm,amssymb} 
\usepackage{color}                  
\usepackage{ulem}  %\sout{......}
                
\begin{document}

\preprint{ECTP-2015-04}    
\preprint{WLCAPP-2015-04}

\title{Possible interrelations among chemical freezeout conditions}

\author{A.~Tawfik\footnote{http://atawfik.net/}}
\affiliation{Egyptian Center for Theoretical Physics (ECTP), Modern University for Technology and Information (MTI), 11571 Cairo, Egypt}
\affiliation{World Laboratory for Cosmology And Particle Physics (WLCAPP), Cairo, Egypt}

\author{ M.~Y.~El-Bakry}
\affiliation{Ain Shams University, Faculty of Education, Department of Physics, Roxi, Cairo, Egypt}

\author{ D.~M.~Habashy}
\affiliation{Ain Shams University, Faculty of Education, Department of Physics, Roxi, Cairo, Egypt}

\author{ M.~T.~Mohamed}
\affiliation{Ain Shams University, Faculty of Education, Department of Physics, Roxi, Cairo, Egypt}

\author{E.~Abbas}
\affiliation{World Laboratory for Cosmology And Particle Physics (WLCAPP), Cairo, Egypt}
\affiliation{Ain Shams University, Faculty of Education, Department of Physics, Roxi, Cairo, Egypt}

\date{\today}

\begin{abstract}

At thermal equilibrium, different chemical freezeout conditions have been proposed so far. They have an ultimate aim of proposing a universal description for the chemical freezeout parameters ($T_{ch}$ and $\mu_b$), which are to be extracted from the statistical fitting of different particle ratios measured at various collision energies with calculations from thermal models. A systematic comparison between these conditions is presented. The physical meaning of each of them and their sensitivity to the hadron mass cuts are discussed. Based on availability, some of them are compared with recent lattice calculations. We found that most of these conditions are thermodynamically equivalent, especially at small baryon chemical potential. We propose that further crucial consistency tests should be performed at low energies. The fireball thermodynamics is another way of guessing conditions describing the chemical freezeout parameters extracted from high-energy experiments. We endorse the possibility that the various chemical freezeout conditions should be interpreted as different aspects of one universal condition.
    
\end{abstract}. 

\pacs{24.10.Pa,12.38.Mh,05.70.Ce}
\keywords{Heavy-ion collisions, hadron resonance gas, chemical freezeout conditions,  QCD phase diagram}
\maketitle

\section{Introduction}

The high temperature and/or large energy density which are likely in relativistic collisions of nuclei are conjectured to derive the quantum-chromodunamic (QCD) matter, the hadrons, to deconfined quarks and gluons. The new phase of matter, the quark-gluon plasma (QGP), has been produced in the relativistic heavy-ion collider (RHIC) \cite{mG}. Reducing the QGP temperature disposes hadronization and the formed fireball is conjectured to go into chemical equilibrium, where the particle abundances are assumed to get fixed. Studying QGP and phase transition(s) at different collision energies belong to the main goals of the heavy-ions collisions experiments. This is partly possible through mapping out the QCD boundary separating hadrons from QGP phase.  At very small baryon chemical potential ($\mu_b$), the hadron-quark phase transition takes place as smooth cross-over \cite{crossover}, where the value of the pseudo-critical temperature reaches $\sim 155 \pm 9~$MeV for three quark flavors, for instance \cite{LQCD1,LQCD2,Pexpand,effmod1,effmod2}. At large $\mu_b$, the phase transition is of first order and can only be studied in QCD-like models \cite{Tawfik:2014eba,Tawfik:2014uka,Tawfik:2014gga,Ezzelarab:2015tya,Tawfik:2015apa}.
        
The hadron resonance gas (HRG) model~\cite{Tawfik:2014eba,HRG1,HRG2a} is utilized by many authors in order to extract information about the chemical freezeout stage, in the final state. Confronting HRG calculations to the experimentally measured particle ratios or yields is implemented to find out the equilibrium chemical freezeout parameters (temperature $T_{ch}$ and corresponding baryon chemical potential $\mu_b$) assuring best fits. Various sets of these parameters are collected over the last $2-3$ decades in the Large Hadron Collider (LHC), RHIC, the Super Proton Synchrotron (SPS), the Alternating Gradient Synchrotron (AGS) and the Schwerionen Synchrotron (SIS) facilities~\cite{Andronic2006,cleymans2006,Tawfik2013,Stachel2014}. It was found that they follow regular patterns with the nucleus-nucleus center-of-mass energy ($\sqrt{s_{NN}}$). It was found that the baryon chemical potential ($\mu_b$) and  $\sqrt{s_{NN}}$ are indirect proportional to each other. At LHC, $\mu_b$ becomes very small indicating a nearly vanishing net-baryon density  \cite{Tawfik:2014eba}. On the other hand, the freezeout temperature increases with increasing $\sqrt{s_{NN}}$. At high energies, $T_{ch}$ reaches a limiting temperature, which apparently sets on at $\sqrt{s_{NN}}>30~$GeV \cite{Andronic2006,Tawfik2013}. Proposing a universal condition describing the dependence of  $T_{ch}$ on $\mu_b$ or on $\sqrt{s_{NN}}$ inters the literature under the name {\it "conditions for chemical freezeout"}.

At equilibrium, the chemical freezeout boundary can be determined from different universal conditions \cite{cleymans1998,Braun-Munzinger2002,Satz2003,Tawfik2006A,Tawfik2006B,Tiwari2012,Oliinychenko2013,Tawfik:2013pd,Tawfik2013trace}. A first condition was proposed in Ref.~\cite{cleymans1998}. It assumes that $T_{ch}$ and $\mu_b$ are characterized by a constant average energy per particle $\sim 1~$GeV~\cite{cleymans1998}. Other phenomenological conditions have been proposed to give further explanations for the extracted $T_{ch}$ and $\mu_b$ parameters at various energies. Their short list is chronically ordered as follows.
\begin{itemize}
\item Sum of the baryon and antibaryon densities $n_{b}+n_{\bar b} = 0.12~$fm$^{-3}$~\cite{Braun-Munzinger2002}.
\item Condition based on percolation theory \cite{Satz2003}.
\item Normalized entropy density  $s/T^3 \simeq 7$ \cite{Tawfik2006A,Tawfik2006B}.
\item Entropy per particle $s/n = 7.18 $ \cite{Tiwari2012,Oliinychenko2013,Tawfik:2013pd}.
\item Trace anomaly $(\epsilon-3p)/T^4=7/2$, where $p$ is pressure and $\epsilon$ is energy density \cite{Tawfik2013trace}.
\item $\kappa\, \sigma^2 = 0$  where $\sigma^2$ and  $\kappa$ are variance and kurtosis of conserved net charge respectively~\cite{Tawfik2013ksigma}.
\item Equal collision and expansion time \cite{Blaschke}.
\end{itemize}
Except for the last two conditions and the percolation theory, the remaining conditions shall be discussed in the present work.

It is obvious that the condition $s/T^3=7$ seems to coincide with $\epsilon/n=1.08~GeV$~\cite{cleymans2006}. Furthermore, at high $\sqrt{s_{NN}}$ (small $\mu_b$), the dimensionless conditions $s/n$ and $s/T^3$ are found almost equal, both approximative are $7$.  This observation raises the question if there are interrelations among the various chemical freezeout conditions? A comparison between three chemical freezeout conditions have been consider about ten years ago \cite{cleymans2006}. In the present, we extend it to include additional chemical freezeout conditions and to consider recent RHIC and LHC results. The main aspect of present work is finding out interrelations between some of the so-far proposed freezeout conditions.

In the present work, we study the equilibrium freezeout criteria and discuss their physical meanings. We also investigate their validity or ability in describing the recently deduced $T_{ch}$ and $\mu_b$ results. Upon availability, some conditions are confronted to recent lattice QCD calculations~\cite{LQCDEoS}. The possible interrelations among different freezeout conditions are analysed, thermodynamically. It is found that many chemical freezeout conditions are thermodynamically equivalent, especially at small baryon chemical potential.  The fireball thermodynamics is another way of guessing conditions against equations using $T_{ch}$ and $\mu_b$ as extracted from particle ratios. To this end, the energy dependence of $T_{ch}$ and $\mu_b$ shall be utilized in estimating the fireball thermodynamic properties at the chemical freezeout.

The present paper is organized as follows. Section \ref{sec:hrg} elaborates details about the hadron resonance gas (HRG) model. 
The different chemical freezeout conditions shall be discussed in section \ref{sec:CFOC}. The possible interrelations among the various chemical freeze-out conditions shall be discussed in section \ref{sec:connect}. Section \ref{sec:fireball} shall be devoted to the properties of fireball thermodynamics at chemical freeze-out. the final conclusions and outlook shall be outlined in section \ref{sec:con}.

\section{Hadron resonance gas model}
\label{sec:hrg}
    
In the HRG model~\cite{Tawfik:2014eba,HRG1,HRG2a}, the formation of hadrons is controlled by the phase space and the conservation laws. For a large number of produced particles, the implementation of grand canonical ensemble (GCE) treatment is obviously justified. Assuming that the strong interactions are dominated by the dynamics leading to the resonance formation, these interactions are believed to be taken into consideration through including heavy resonances~\cite{interaction1,interaction2}. In the hadronic phase, the grand canonical partition function is summed up over {\it single-particle partition functions} of stable hadron and unstable resonances 
\begin{equation}
\ln Z(T, \mu ,V)=\sum_i\pm \frac{V g_i}{2\pi^2}\int_0^{\infty}p^2\,   \ln\left\{1\pm \lambda_i\exp\left[\frac{-\varepsilon_i(p)}{T}\right]\right\}\, dp, \label{eq:lnz}
\end{equation} 
where $\lambda_i = \exp(\mu_i/T)$ is the fugacity of $i$-th hadron, $\varepsilon_i(p)=(p^2+ m_i^2)^{1/2}$ is its dispersion relation, $g_i$  is the degeneracy factor and $\pm$ stands for fermions and bosons, respectively. The  chemical potential related to $i$-th particle is given as \hbox{$\mu_i=\mu_b B_i+\mu_s S_i+\mu_{q} Q_i$}, where $ \mu_s$, $\mu_q$, $B_i$, $S_i $ and $Q_i$ are strange, and charge chemical potential and baryon, strange and charge quantum number, respectively. 

The thermodynamic properties of this QCD system can be obtained from the partition function, Eq. (\ref{eq:lnz}). In non-degenerate case, the pressure and number density of the $i$-th hadron or resonance can be written as
\begin{eqnarray} 
P_i(T,\mu_i ) &=& \pm\frac{ g_i T^2 m_i^2}{2\pi^2}\sum_{n=1}^\infty\frac{\left(\pm\lambda_i\right)^{n}}{ n^2} K_2\left({nm_i\over T}\right), \label{eq:pres} \\
n_i(T, \mu) &=& \pm\frac{g_i T m_i^2}{2\pi^2}\sum_{n=1}^\infty\frac{\left(\pm\lambda_i\right)^{n}}{n} K_2\left({nm_i\over T}\right). \label{eq:n_i}
\end{eqnarray}
where here $\mp$ stands for fermions and bosons, respectively. The chemical potential and temperature are there to assure the conservation of strangeness, charge and baryon numbers. This can be fulfilled by bearing off the volume as \hbox{$\sum_i\ n_i(T, \mu_i) S_i =0$},  
\hbox{$\sum_i n_i(T,\mu_i) Q_i/\sum_i n_i(T,\mu_i) B_i=Z/A$}, where $A$ and $Z$ are the mass and atomic number of the colliding nuclei.
So far, all expressions are quantum statistical. In discussing the results, the logarithm series is extracted, where the first term in Eqs. \eqref{eq:pres} and \eqref{eq:n_i} stand for Maxwell-Boltzmann (MB) limit. We remark that the calculations of all quantities are performed through quantum statistical expressions.  
 
It is worthwhile to highlight that the summation in the previous expressions includes contributions from light flavor hadrons. These are listed in the particle data group (PDG)~\cite{PDG} with zero-width approximation. Corrections due to repulsive interaction between hadrons at small distances (known as excluded-volume correction (EVC)~\cite{EVC}) is not taken into account, as it was concluded that these are practically irrelevant \cite{Tawfik2013trace,Tawfik2013ksigma}. Furthermore, the decay of unstable resonances is not taken into consideration. 

In this regard, it intends to study the properties of the fireball at the stage of the chemical freezeout. It is worthwhile to mention that the calculations from the HRG model are performed at full chemical equilibrium, as it was evidenced that in most central high-energy collisions the system reaches full chemical equilibrium \cite{Tawfik:2014dha}.

\section{Chemical freezeout conditions}
\label{sec:CFOC}

It is conjectured that the quark-hadron phase transition occurs prior to the stage of  chemical freezeout, which is defined as the state at which the inelastic interactions between produced hadrons cease as the mean free path becomes larger than the system size \cite{Cleymans1993}. Consequently, the hadron yields are frozen, i.e. no further change in the particle number (chemistry is fixed) takes place. Fitting particle ratios (to maximally eliminate the volume effect) calculated to the thermal models to the experimentally measured ones results in the freezeout parameters, which are interpreted in terms of some universal conditions 
~\cite{cleymans1998,Braun-Munzinger2002,Satz2003,Tawfik2006A,Tawfik2006B,Tiwari2012,Oliinychenko2013,Tawfik:2013pd,Tawfik2013trace}. It is in order now to introduce their basic concepts and compare between their predications, especially in reproducing the recent experimental results on particle ratios and yields.

\subsection{Average energy per particle}

This criterion was proposed in Ref.~\cite{cleymans1998} as a condition describing the dependence of $T_{ch}$ on $\mu_b$ as extracted from SPS, AGS and SIS facilities. Later on, other parameters extracted at top SPS and RHIC energies match well with this condition~\cite{Tawfik:2014eba,HRG1}. In a non-relativistic system, i.e. MB limit, the average energy per particle for $i$-th hadron or resonance reads \cite{Cleymans1999},
\begin{equation} 
\frac{\epsilon}{n}  = \sum_i m_i \left[1 + \frac{3}{2}\frac{T}{m_i} +\frac{15}{8}\left(\frac{T}{m_i}\right)^2+\cdots\right]. \label{eq:CFOC1} 
\end{equation}
\begin{enumerate}
\item At low energies, it is conjectured that the system is dominated by nucleons. By using the nucleons mass $\approx 0.94$ GeV, $\epsilon/n$ will have a little bit higher value \cite{Cleymans1999}. 
\item At high energies, the system will be dominant by light mesons where $\epsilon/n$ can be considered as an average particle mass plus the temperature terms.  Although the average mass of the particles in the system decreases, the temperature will increase~\cite{Cleymans1999}.
\end{enumerate}
Joining all chemical freezeout parameters through $\epsilon/n\simeq1~$GeV, seems to be a good guess for a universal freezeout condition \cite{Cleymans1999}.

It is worthwhile to recall that by assumed that the multi-hadron production is the QCD counterpart of the Hawking-Unruh radiation \cite{Castorina2007}, it was predicted that the hadronic freezeout from the black-hole radiation for the average energy per hadron gives $1.09~$GeV \cite{Castorina2014,Tawfik:2015fda}. This value amazingly agree with the thermal models. Thus, both approaches; namely thermal models and black hole radiation, not only confirm the idea of chemical freezeout condition but also approximately estimate the same value of averaged energy per particle at finite baryon chemical potential \cite{Tawfik:2015fda}.

\begin{figure} 
\begin{center} 
\includegraphics[scale=0.5]{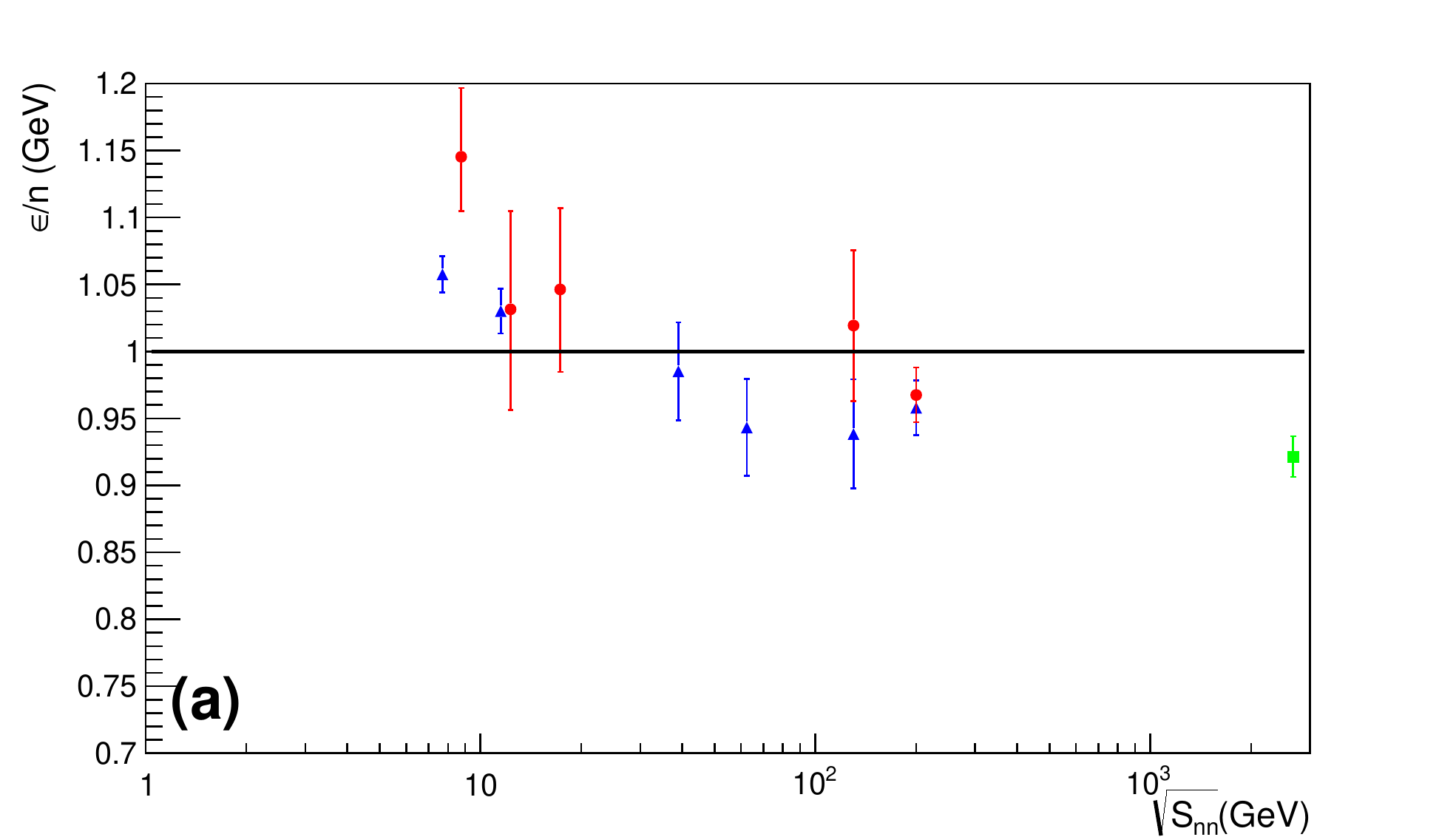} 
\caption{\footnotesize The dependence of $\epsilon/n$ on energy at the freezeout parameters $\mu_b$ and $T_{ch}$  ~\cite{Andronic2006,Tawfik2013,Stachel2014} is depicted. The closed triangle symbols are results from Ref.~\cite{Tawfik2013}, the circles stand for results from Ref.~\cite{Andronic2006} and the square are the results at LHC energy \cite{Stachel2014}. The solid line represents $\epsilon/n=1~$GeV as calculated from HRG model.} 
\label{fig:en_massL}  
\end{center}  
\end{figure}

In order to determine the range of energies (or chemical potentials) in which this freeze-out condition $\epsilon/n=1~$GeV (solid line) is valid, we compare with $\epsilon/n$ calculated from the extracted parameter using GCE and full chemical equilibrium~\cite{Andronic2006,Tawfik2013,Stachel2014} at different energies. The comparison is depicted in  Fig. \ref{fig:en_massL}. The closed triangle symbols are results from Ref.~\cite{Tawfik2013}, the circles show the results from Ref.~\cite{Andronic2006} and the squares stand for results at LHC energy \cite{Stachel2014}. It is obvious that $\epsilon/n=1~$GeV (solid line) lays above the phenomenologically deduced results, especially at the LHC energy. This is a natural consequence of determining $T_{ch}$ at the ultra-relativistic high-energies $2.67~$TeV, which is relatively less than the one observed at top RHIC energies. Concretely, in GCE and assuming full chemical equilibrium, $\epsilon/n$ is calculated by using the same parameters, $T$ and $\mu_b$, which have been presented in Refs.~\cite{Andronic2006,Tawfik2013,Stachel2014} at different energies.

\begin{figure} 
\begin{center} 
\includegraphics[scale=0.4]{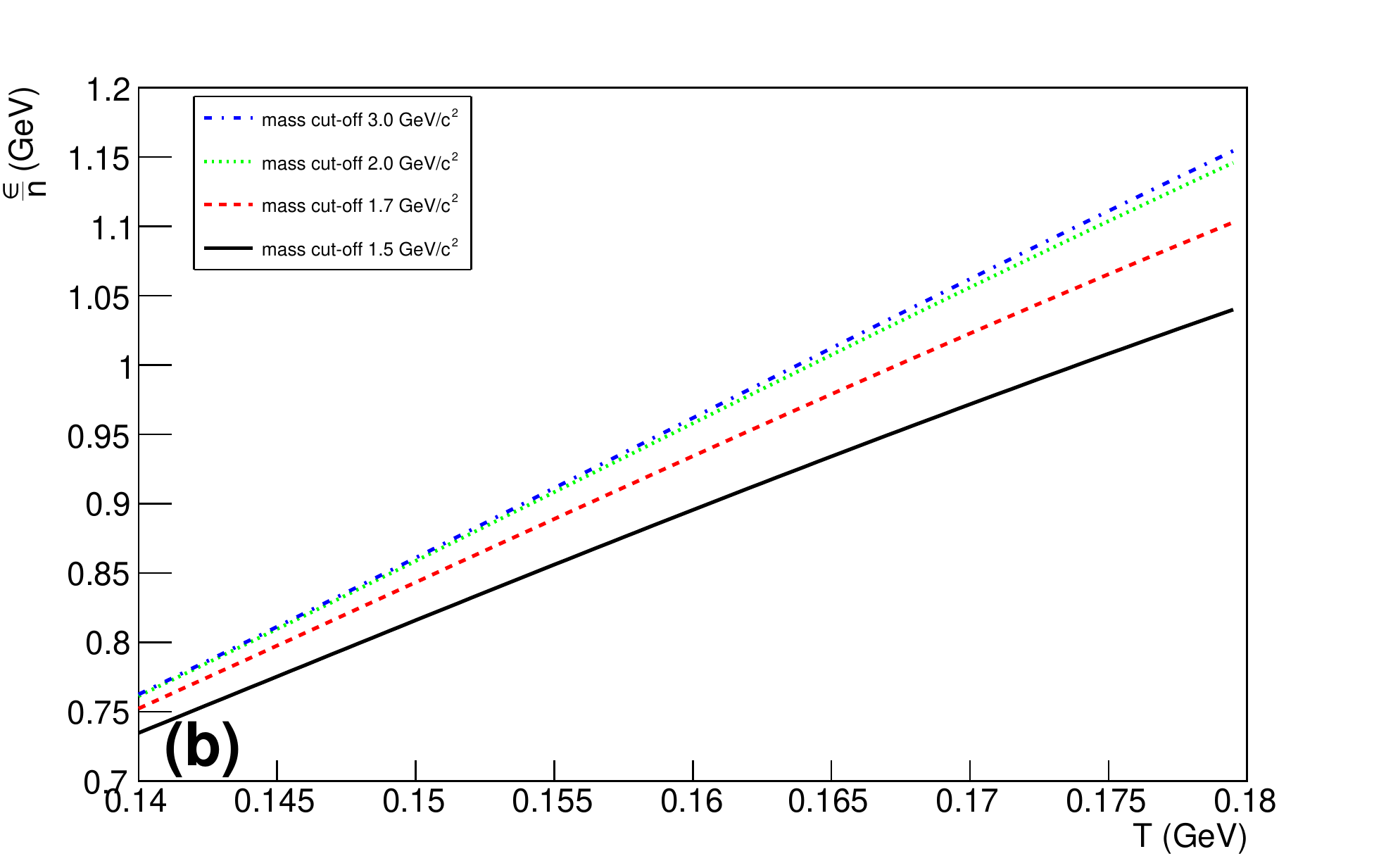}
\includegraphics[scale=0.43]{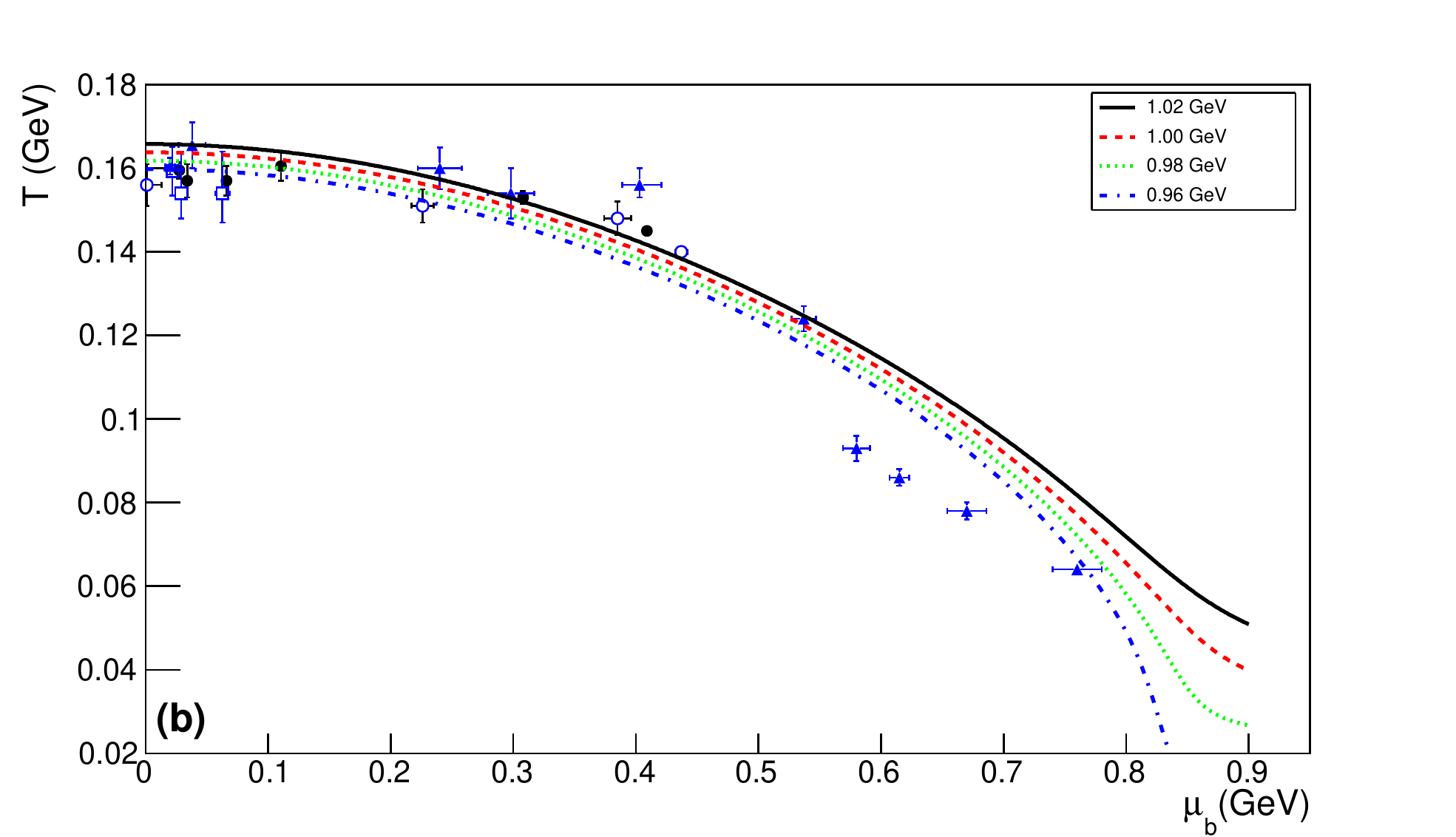}
\caption{\footnotesize Left-hand panel (b) shows the extracted parameters $T_{ch}$ and $\mu_b$ at different values of  $\epsilon/n$. The closed diamond symbols are results from Ref.~\cite{Andronic2006}, the open circles sand for results from Ref.~\cite{Becattini2013,Stock2013}, the open squares are the STAR results at $0-5\%$ centrality ~\cite{STAR2} and the closed circles represent our results~\cite{Tawfik2013}. Right-hand panel (c) shows the dependence of $\epsilon/n$ on temperature at different resonance mass cut-off.  } 
\label{fig:en_mass}  
\end{center}  
\end{figure} 
  
In left-hand panel of Fig.\ref{fig:en_mass} (b), the results on freezeout parameters calculated from HRG model under the condition of constant averaged energy per particle, are confronted to $\mu_b$ and $T_{ch}$, which are deduced from statistical fitting of various experimental particle ratios and corresponding calculations from HRG model, in which  $\mu_b$ and $T_{ch}$ are free parameters \cite{Andronic2006,Tawfik2013,Becattini2013}. It is apparent that the most suitable values of $\epsilon/n$ are the ones less than $1~$GeV whose  limiting temperatures  also agree well with the pseudo-critical temperature range predicted by the lattice QCD calculations~\cite{LQCD1,LQCD2}. At high energies, a good agreement between this freezeout condition and the extracted parameters is found. The calculated values are a little bit higher than the phenomenologically deduced ones, especially at low energy. It should be stressed here that these calculations have been preformed under GCE with zero-width which is not valid at AGS and SIS energies.

At vanishing baryon chemical potential, the right-hand panel Fig.\ref{fig:en_mass} (c) illustrates the dependence of $\epsilon/n$ on the freezeout temperature at various limits for mass cut-off, i.e. including/removing hadron resonances.  It is obvious that the freezeout condition is obviously effected by the mass cut. With increasing the mass limits, the resonance contributions increase and consequently $\epsilon/n$ seems to increase as well. At different hadron resonance masses, different $T_{ch}$ are able to fulfil the condition of constant $\epsilon/n$. It has been assumed that EVC, in which a single hard-core is assumed for all hadrons and resonances in the MB limit has the same effect on both $n$ and $\epsilon$~\cite{cleymans2006}. Both quantities shall be suppressed by almost the same factor~\cite{EVC}.

\subsection{Baryon and antibaryon particle density}

The total baryon number density, $n_{b}+n_{\bar b}$, was proposed as a condition interpreting the extracted freeze-out parameters  \cite{Braun-Munzinger2002}. Relative to the meaning of constant energy per particle, here another interpretation for the freeze-out parameters at different collision energies is proposed \cite{Braun-Munzinger2002}. The correlations of baryons, such as baryon-baryon and baryon-meson interactions, are assumed to be the processes responsible for the chemical equilibrium~\cite{Braun-Munzinger2002}. The existing of such interactions weakens the applicability of this condition, as the chemical freeze-out is defined as a stage in which elastic scatterings become dominant, while this freezeout condition apparently relies on inelastic interactions, which likely drive the system towards chemical nonequilibrium. 

In MB limit, the freezeout condition of the baryon density can be expressed as
\begin{equation} 
\label{eq:CFOC2} 
 n_{b}+n_{\bar b} =  \sum_i \frac{g_i\, T\, m_i^2}{2\, \pi^2} K_2\left(\frac{m_i}{T}\right)~\cosh\left(\frac{\mu_i}{T}\right), %\qquad\qquad
%\cosh\left(\frac{\mu_i}{T}\right) = 1+ \frac{1}{2!} \left(\frac{\mu_i}{T}\right)^{2} +  \frac{1}{4!}\left(\frac{\mu_i}{T}\right)^{4}+\cdots,
\end{equation}
where $n_b$ is the baryon number density. 

\begin{figure} 
\begin{center} 
\includegraphics[scale=0.5]{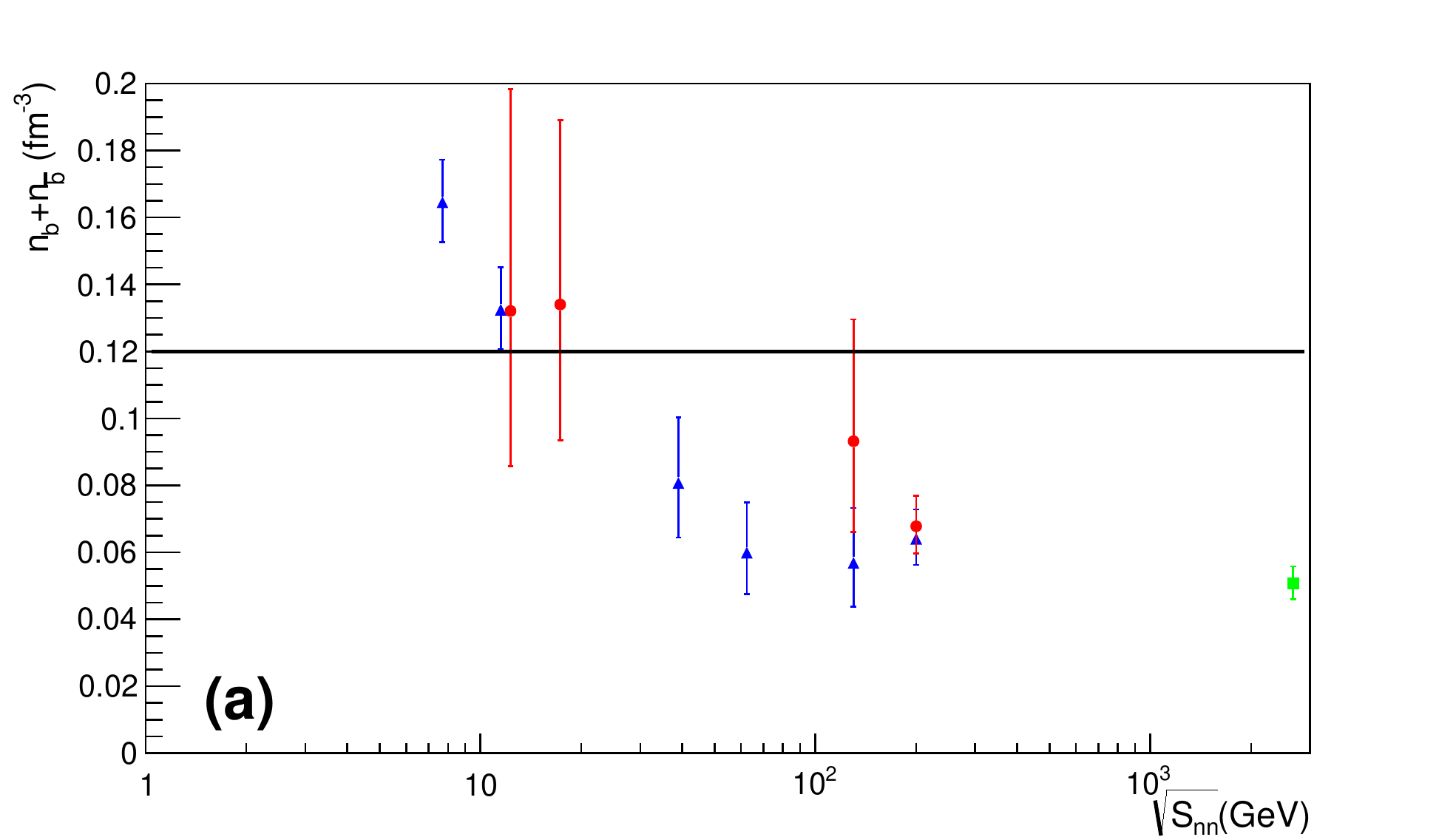}
\caption{\footnotesize The same as in Fig. \ref{fig:en_massL} but here for $n_{b}+n_{\bar b}$. 
%on energy is calculated at the freezeout parameters $\mu_b$ and $T_{ch}$  ~\cite{Andronic2006,Tawfik2013,Stachel2014}. The closed triangle symbols are results from Ref.~\cite{Tawfik2013}, the circles stand for results from Ref.~\cite{Andronic2006} and the square is the results at LHC energy \cite{Stachel2014}. 
The solid line represents the constant value  $n_{b}+n_{\bar b} = 0.12 fm^{-3}$. } 
\label{fig:nbbar_massL}  
\end{center}  
\end{figure}

\begin{figure} 
\begin{center} 
\includegraphics[scale=0.4]{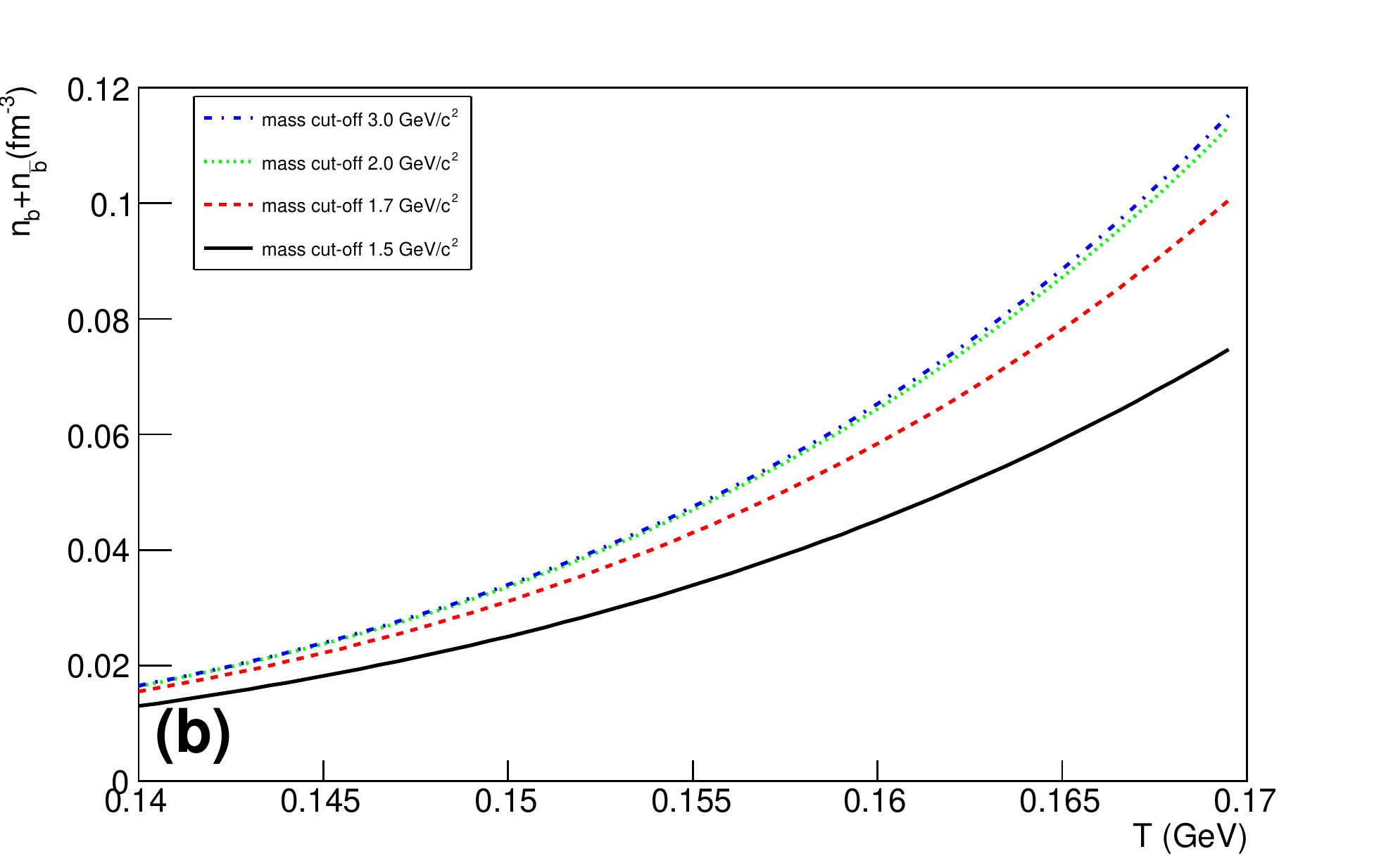}
\includegraphics[scale=0.43]{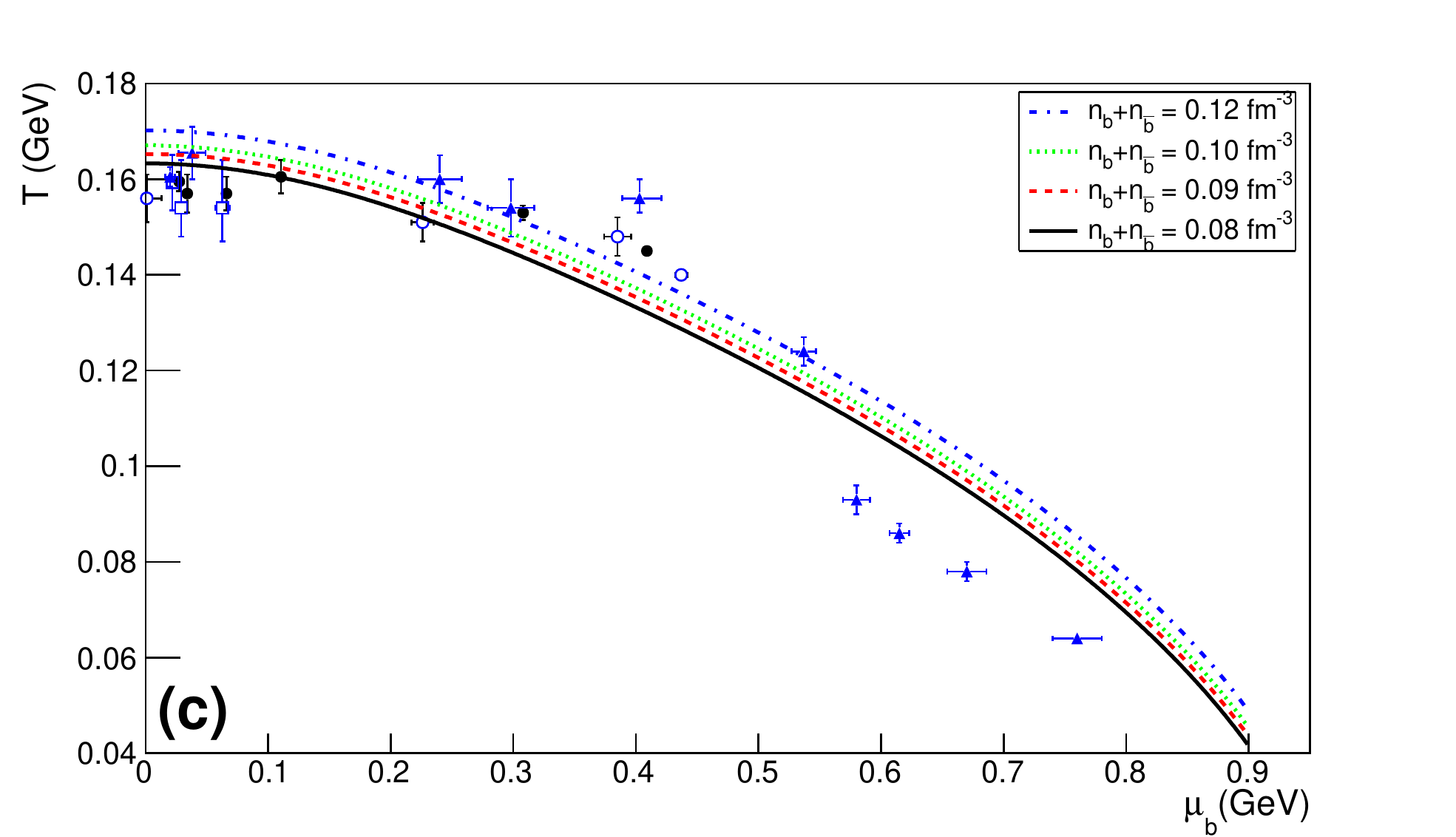}
\caption{\footnotesize Left-hand panel (b) shows the dependence of $n_{b}+n_{\bar b}$ on temperature at different resonance  masses. Right-hand panel (c) presents different values for $n_{b}+n_{\bar b}$ and accordingly the extracted parameters $T_{ch}$ and $\mu_b$. The closed diamond symbols are results from Ref.~\cite{Andronic2006}, the open circles give the results from Ref~\cite{Becattini2013}, the open squares are STAR results at $0-5\%$ centrality ~\cite{STAR2} and the closed circles represent our results~\cite{Tawfik2013}. } 
\label{fig:nbbar_mass}  
\end{center}  
\end{figure}

\begin{enumerate}
\item At very high energies, i.e. $\mu_i/T<1$, the quantity $n_{b}+n_{\bar{b}}$ remains fixed. In this energy limit, $T_{ch}$ remains constant, as well, while $\mu_b$ decreases with the energy. 
\item At low energies, this criteria is effected by two factors; the increase in the hyperbolic function and the decrease in the other functions with increasing $\mu_i$ and decreasing temperature.  
\item In limit of $T \rightarrow 0$, the value of $n_{b}+n_{\bar{b}}$ will be nearly the nucleus baryon density.
\end{enumerate}

We compare $n_{b}+n_{\bar{b}}$ with the calculations from GCE at full chemical equilibrium~\cite{Andronic2006,Tawfik2013,Stachel2014} and different energies. The comparison is depicted in  Fig.  \ref{fig:nbbar_massL}. It is obvious that $n_{b}+n_{\bar{b}}$ increases at low energy. This could be due to appearance of baryons at mid-rapidity. At high energy, e.g. $>40~$GeV, this freezeout condition takes a constant value because the change in $\mu_b/T$ becomes no longer significant.

This freeze-out condition is also effected by the hadron mass cuts. In Eq. \ref{eq:CFOC2}, if the masses of included hadron resonances increase, the number of terms that shall be summed up to each others raises and consequently  $n_{b}+n_{\bar{b}} $ increases. Also, we notice that the different values of $T_{ch}$ are able to fulfil the condition at different masses of hadron resonances as shown in left-hand panel of Fig. \ref{fig:nbbar_mass} (b). It is obvious that $n_{b}+n_{\bar{b}}$ shall be suppressed when taking EVC into consideration~\cite{EVC}. Here, the suppression factor is not the same at all energies even when the hard-core radius is fixed. Thus, it is very essential to estimate the effects of EVC when implementing this freezeout condition. As mentioned earlier, EVC is not applied in the present calculations.
       
The right-hand panel of Fig. \ref{fig:nbbar_mass} (c) shows $T_{ch}$ vs. $\mu_b$ at different values of  $n_{b}+n_{\bar{b}}$ compared with the chemical freeze-out parameters $T_{ch}$ and  $\mu_b$, which are extracted from the fits of measured particle ratios and the HRG calculations, in which $T_{ch}$ and  $\mu_b$ are taken as free parameters \cite{Andronic2006,Tawfik2013,Becattini2013}. The parameters extracted at constant  $n_{b}+n_{\bar{b}}$ show that at very high energies $n_{b}+n_{\bar{b}}$ should be reduced to $0.08~$fm$^{-3}$. We notice that  $n_{b}+n_{\bar{b}}$ increases with decreasing the energy (or increasing the chemical freezeout).

\subsection{Normalized entropy density}

It has been argued that the chemical freeze-out parameters can be described by a constant normalized entropy density (entropy density divided by $T^3$)~\cite{Tawfik2006A,Tawfik2006B}. The quantity $s/T^3$ is assumed to measure the degrees of freedom (dof). The constant value refers to constant dof in the hadronic phase. During the phase transition the hadrons' dof should be replaced by the QGP's ones. Concretely, the value of $s/T^3$ was chosen at the pseudo-critical temperature as calculated in the lattice QCD simulations \cite{Karsch:2003vd,Karsch:2003zq,Redlich:2004gp,Tawfik:2004sw}. This assigned value is compatible with the quark flavors and masses used in the lattice calculations at vanishing baryon chemical potential. It was found that $s/T^3=5$ for two quark flavors and $s/T^3=7$ for three quark flavors \cite{Karsch:2003vd,Karsch:2003zq,Redlich:2004gp,Tawfik:2004sw}.  Constant $s/T^3$ is conjectured to remain unchanged with increasing $\mu_b$~\cite{Tawfik2006A,Tawfik2006B}. Furthermore, the normalized entropy density was also used to separate a meson-dominant region from baryon-dominant one~\cite{Cleymans2005}. Accordingly, an explanation for the rapid variations of certain particle ratios that was observed at lower SPS energies~\cite{Gazdzicki2004} has been suggested~\cite{Cleymans2005}. It is obvious that when $T \rightarrow 0$, the thermal entropy density vanishes, as well.

\begin{figure} 
\begin{center} 
\includegraphics[scale=0.5]{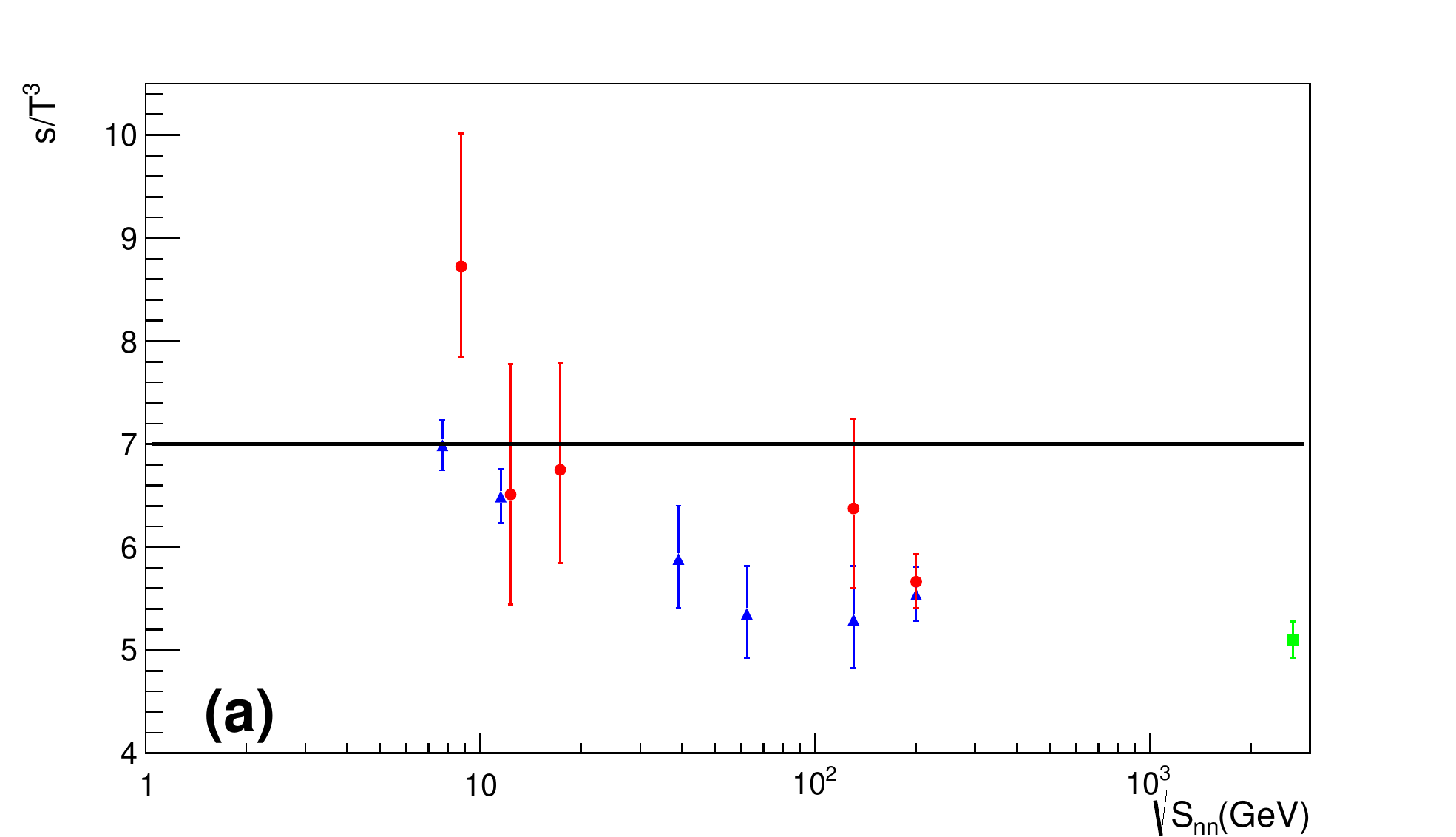}
\caption{\footnotesize The same as in Fig. \ref{fig:en_massL} but here for constant $s/T^3$. 
%on energy calculated at the freezeout parameters $\mu_b$ and $T_{ch}$  ~\cite{Andronic2006,Tawfik2013,Stachel2014}. The closed triangle symbols are results from Ref.~\cite{Tawfik2013}, the circles stand for results from Ref.~\cite{Andronic2006} and the square is the results at LHC energy \cite{Stachel2014}. 
The solid line represents $s/T^3=7$.  } 
\label{fig:ST_massL}
\end{center}  
\end{figure}

\begin{figure} 
\begin{center} 
\includegraphics[scale=0.4]{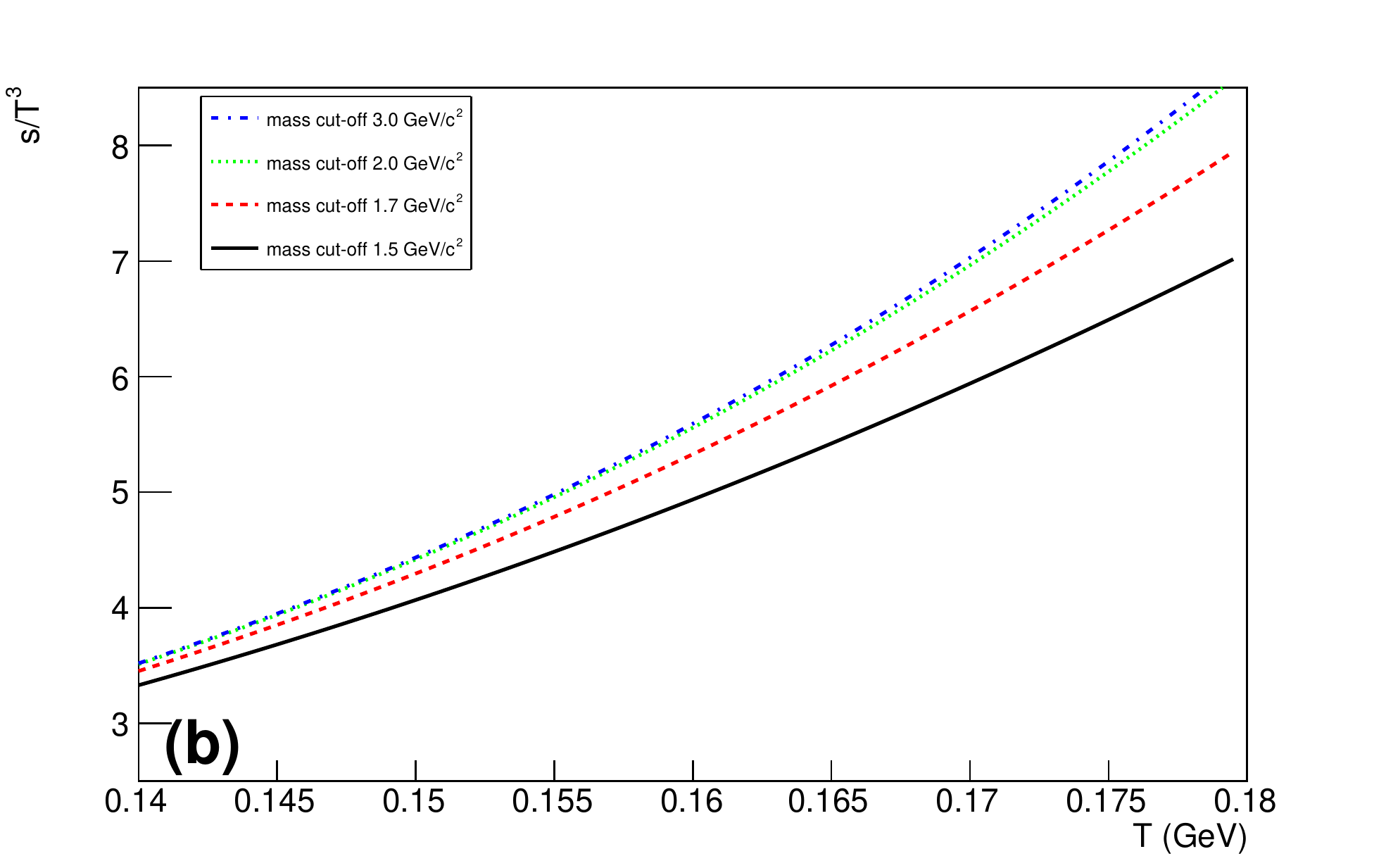}
\includegraphics[scale=0.43]{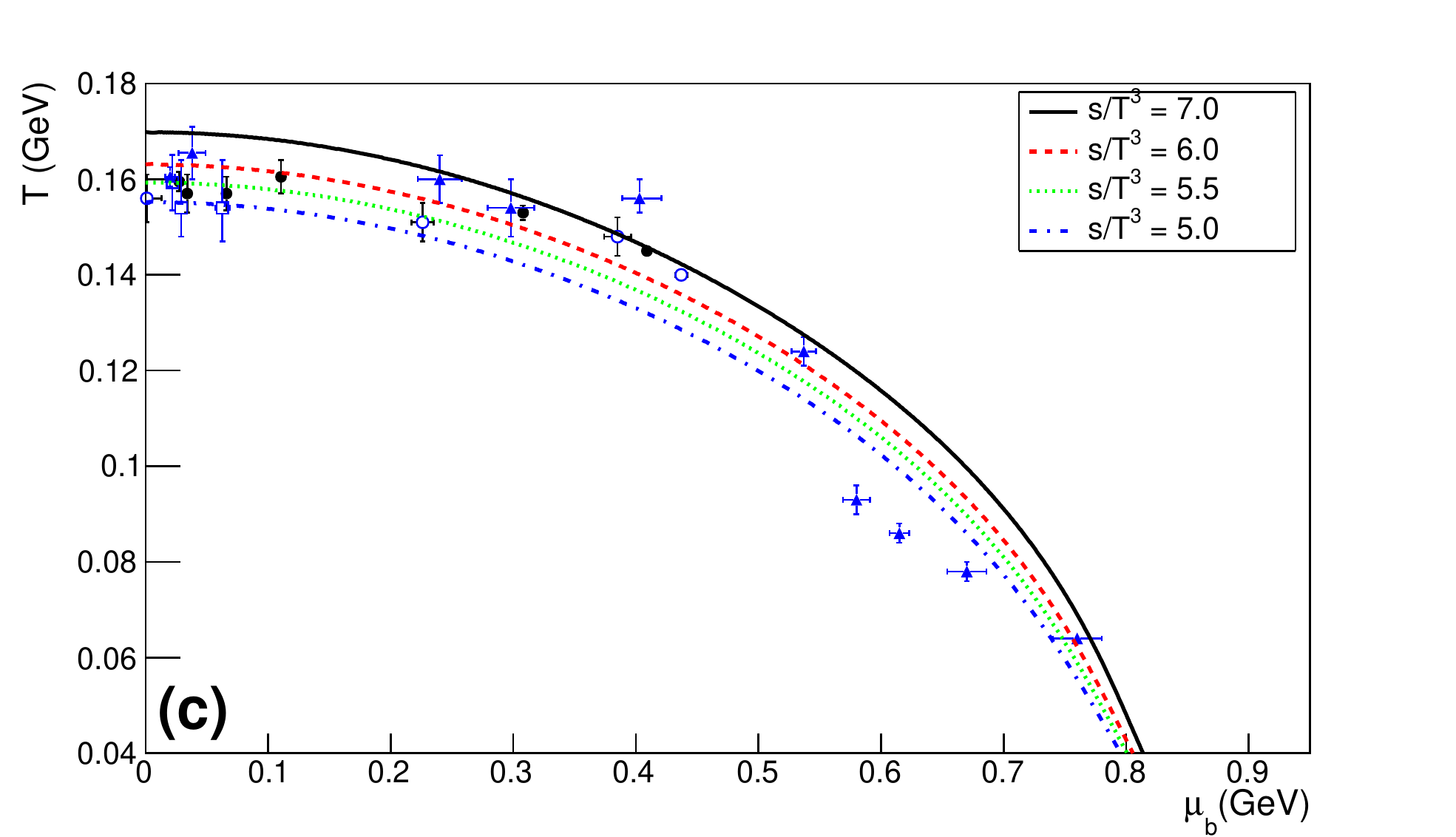}
\caption{\footnotesize Left-hand  panel (b) gives the dependence of $s/T^3$ on the temperature at different resonance masses. Right-hand panel (c) shows different values of $s/T^3$ and the parameters $T_{ch}$ and $\mu_b$ compared with the parameters deduced from the particle ratios. The closed diamond symbols are results from Ref.~\cite{Andronic2006}, the open circles stand for results from Ref.~\cite{Becattini2013}, the open squares are STAR results at $0-5\%$ centrality \cite{STAR2} and the closed circles give our results~\cite{Tawfik2013}. } 
\label{fig:ST_mass}
\end{center}  
\end{figure}

In  Fig. \ref{fig:ST_massL}, $s/T^3$ calculated from the extracted parameter using GCE and full chemical equilibrium~\cite{Andronic2006,Tawfik2013,Stachel2014} is depicted at different energies. $s/T^3 $ increases at low energy (AGS energies) because $T^3$ decreases much faster than $s$ as the energy decreases. Thus, $s/T^3$ becomes no longer constant during chemical freeze-out at very large $\mu_b$. $s/T^3$ value at LHC energy, equals the top RHIC energies values within the error. 

The dof in the hadronic phase depend on the temperature, explicitly. This means that the hadronic dof should not remain fixed, at least from the thermodynamical point-of-view. Also, we highlight that EVC is conjectured to change the value of $s/T^3$ but its physical meaning remains unchanged. In other words, when EVC is implemented one may find another suitable guess for $s/T^3$ which will might less than the proposed values. Even this slightly smaller value does not change the ability of $s/T^3$ to reproduce the freeze-out parameters. At a certain temperature, adding more resonances by increasing the limits of the hadron resonance masses, $s/T^3$ increases as shown in left-hand panel of Fig. \ref{fig:ST_mass} (b). It is found that different values of $T_{ch}$ are able to fulfil the condition of constant $s/T^3$ at different cuts in the resonance masses.
     
The right-hand panel of Fig. \ref{fig:ST_mass} (c)  shows different values of $s/T^3$ and the corresponding parameters $T_{ch}$ and  $\mu_b$ calculated from HRG under the condition of constant $s/T^3$ compared with the ones deduced from fits of measured particle ratios and HRG calculations. We notice that lower values of $s/T^3$ are consistent with the deduced parameters as well as with the lattice QCD calculations. The best agreement is found if $s/T^3$ ranges between $4.70$ and $6.0$, this is corresponding to $T_{ch}$ ranging from $150$ and $160~$MeV \cite{LQCDEoS}.

\subsection{Entropy per particle}

It was assumed that the entropy per particle can be utilized in describing the chemical freeze-out parameters~\cite{Tiwari2012,Oliinychenko2013,Tawfik:2013pd}. The value of $s/n$ was assumed as $\sim 7$. The energy independence of $s/n$ was interpreted as an evidence for the adiabatic chemical hadron production in heavy-ion collisions~\cite{Oliinychenko2013,Tawfik:2013pd}. The entropy per particle is also assumed to measure the average of the available microstates, i.e. similar to the typical meaning of the entropy. Furthermore, $s/n$ was used in framework of nonequilibrium statistical hadronization models in order to explain peaks in some particle ratios at low energy, such as  $K^+/\pi^+$ \cite{Tawfik:2005gk}.

\begin{figure} 
\begin{center} 
\includegraphics[scale=0.5]{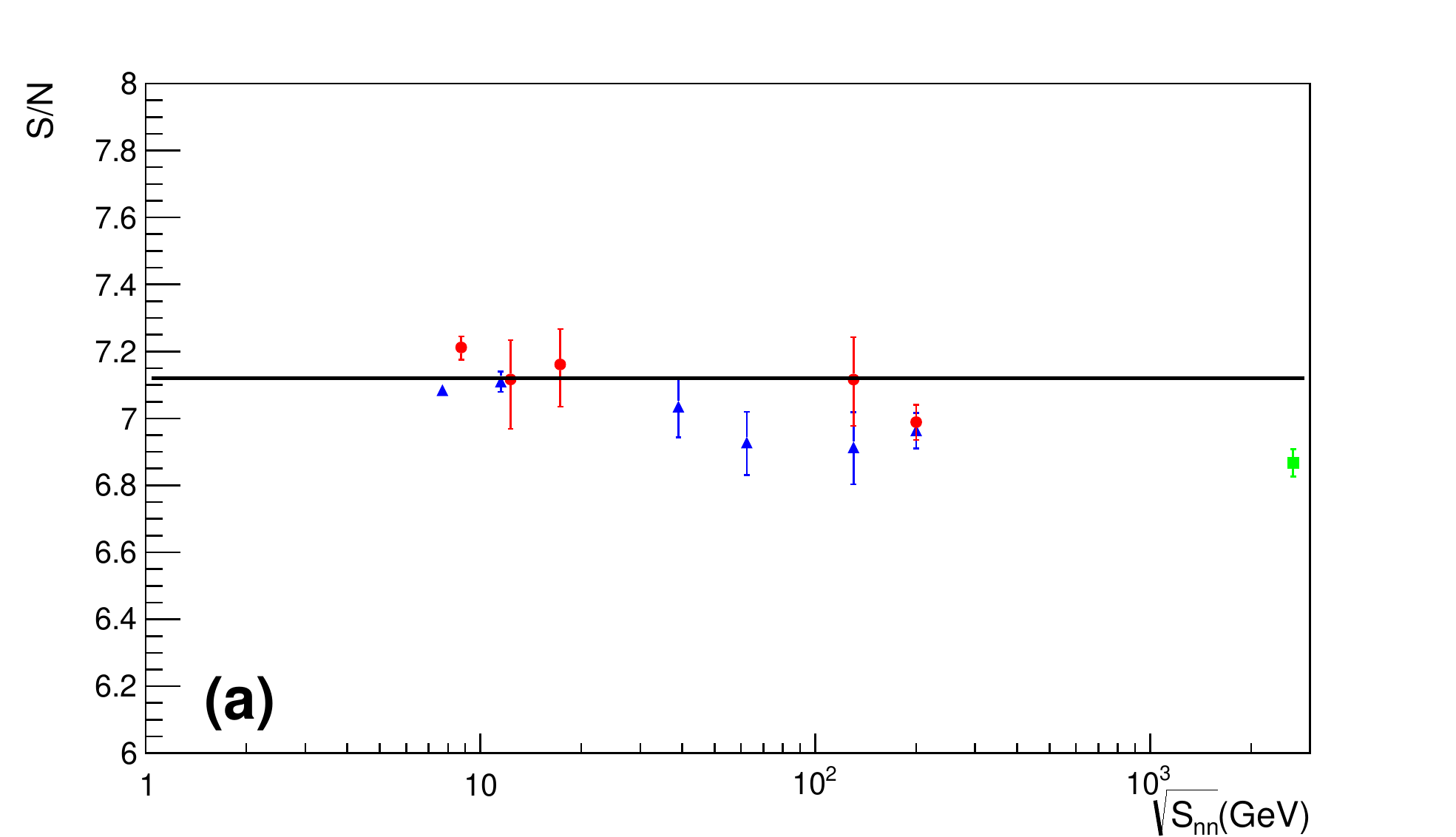}
\caption{\footnotesize The same as in Fig. \ref{fig:en_massL} but here for constant $s/n$. 
% on energy as calculated at the freezeout parameters $\mu_b$ and $T_{ch}$ \cite{Andronic2006,Tawfik2013,Stachel2014}. The closed triangle symbols are results from Ref.~\cite{Tawfik2013}, the circles stand for results from Ref.~\cite{Andronic2006} and the square is the results at LHC energy \cite{Stachel2014}. 
The solid line represents $s/n=7.18$~\cite{Oliinychenko2013,Tawfik:2013pd}. } 
\label{fig:sn_massL}
\end{center}  
\end{figure}

Fig. \ref{fig:sn_massL} shows $s/n$ calculated from the extracted parameter using GCE and full chemical equilibrium~\cite{Andronic2006,Tawfik2013,Stachel2014} at different energies. At high energies, $s/n\simeq 7$ shows a fair constant behaviour. However, $s/n\simeq 7$ requires a very high temperature in order to be satisfied at large baryon chemical potential $ \mu_b>0.5 GeV$. The calculated value at LHC is a lower than the one calculated at RHIC energies. This might be interpreted from the observation that the extracted $T_{ch}$ from the particle ratios at $2.67~$TeV is less than the one at top RHIC energies.
       
\begin{figure} 
\begin{center} 
\includegraphics[scale=0.4]{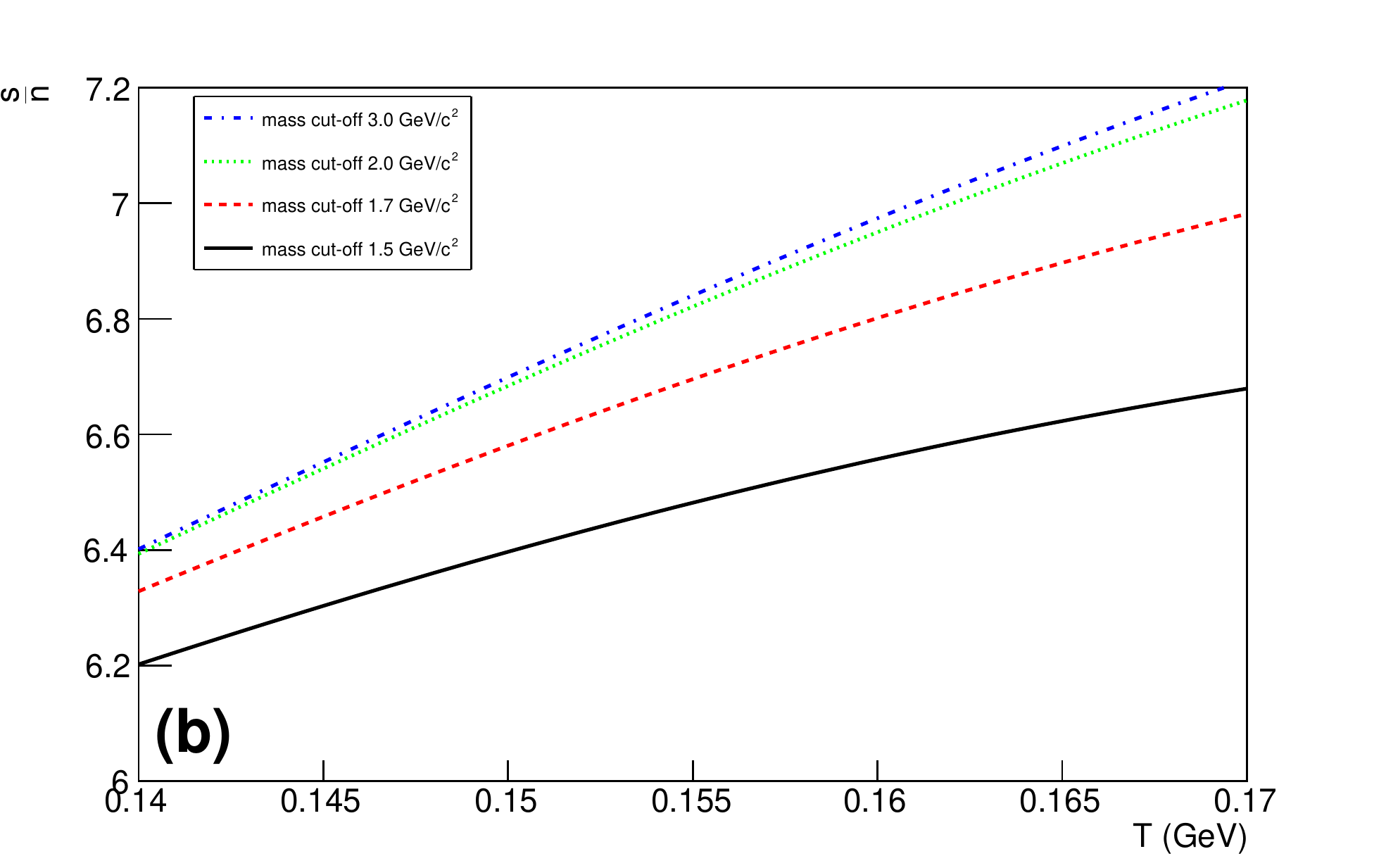}
\includegraphics[scale=0.43]{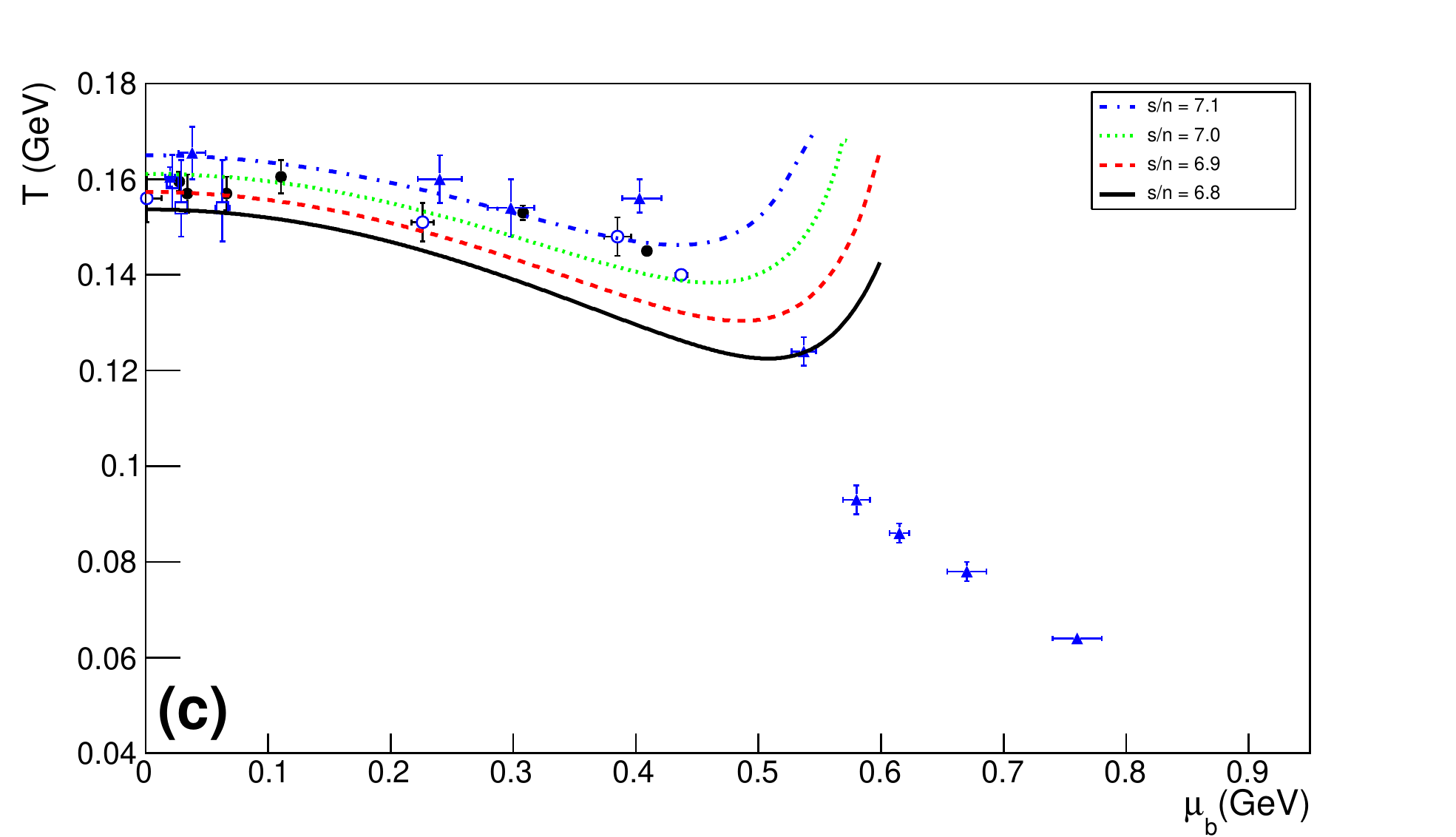}
\caption{\footnotesize Left-hand panel (b) shows the dependence of $s/n$ on temperature at different resonance spectrum masses. Right-hand panel (c) depicts the freeze-out parameters $T_{ch}$ and $\mu_b$ at different values of $s/n$ and compares them with the ones from the particle ratios. The closed diamond symbols are results from Ref.~\cite{Andronic2006}, the open circles stand for the results from Ref. \cite{Becattini2013}, the open squares are the STAR results at $0-5\%$ centrality \cite{STAR2} and the closed circles give our results~\cite{Tawfik2013}. } 
\label{fig:sn_mass}
\end{center}  
\end{figure}

This condition is apparently effected by the hadron mass cuts, especially at vanishing $\mu_b$. A completely different values of $T_{ch}$ can be observed at different cuts of the resonance masses, right-hand panel of Fig. \ref{fig:sn_mass} (b). This also illustrates how $s/n$ is sensitive to the hadron mass cuts and shows that different $T_{ch}$ can be obtained at different values of $s/n$.  If both quantities $n$ and $s$ are suppressed by the same factor, for instance, by assuming a single hard-core for all hadrons in MB limit~\cite{EVC}, it seems safe to study $s/n$ with or without EVC. 

The right-hand panel of Fig. \ref{fig:sn_mass} (c) depicts the freeze-out parameters $T_{ch}$  and  $\mu_b$ deduced at different values of $s/n$ and compared with the recently extracted parameters from fits of experimental particle ratios and their calculations from HRG model. The agreement between the predicted values of $s/n$ and extracted parameters is limited at small $\mu_b$. This suggests a limiting value of  $s/n$  around $7$ at high temperature but when the temperature decreases the extracted parameters likely suggest smaller values, as $s/n$ diverges at large $\mu_b$ \cite{Tawfik:2013pd}. This can be understood as a decrease in the average possible microstates with increasing temperature, which obviously contradicts the second law of thermodynamics.

\subsection{Trace Anomaly}

The QCD trace anomaly $(\epsilon-3p)/T^4$ (also known as interaction measure) is finite in the hadron phase and vanishes in the QGP phase. According to lattice QCD simulations, the trace anomaly shows a peak near the pseudo-critical temperature indicating the phase transition and the appearance of massive quasiparticles \cite{LQCDtrace}. This quantity was used as a novel chemical freeze-out condition \cite{Tawfik2013trace} reproducing a universal description for $T_{ch}$ and $\mu_b$. The value assigned to it is deduced from the lattice QCD calculations. It is worthwhile to highlight that the lattice QCD calculations \cite{LQCDtrace1} show a shift in the trace anomaly curve towards lower $T$ at large $\mu_b$ \cite{LQCDtrace1}. This means that the resulting $T_{ch}$ would be decreasing with increasing $\mu_b$. 

\begin{figure} 
\begin{center} 
\includegraphics[scale=0.5]{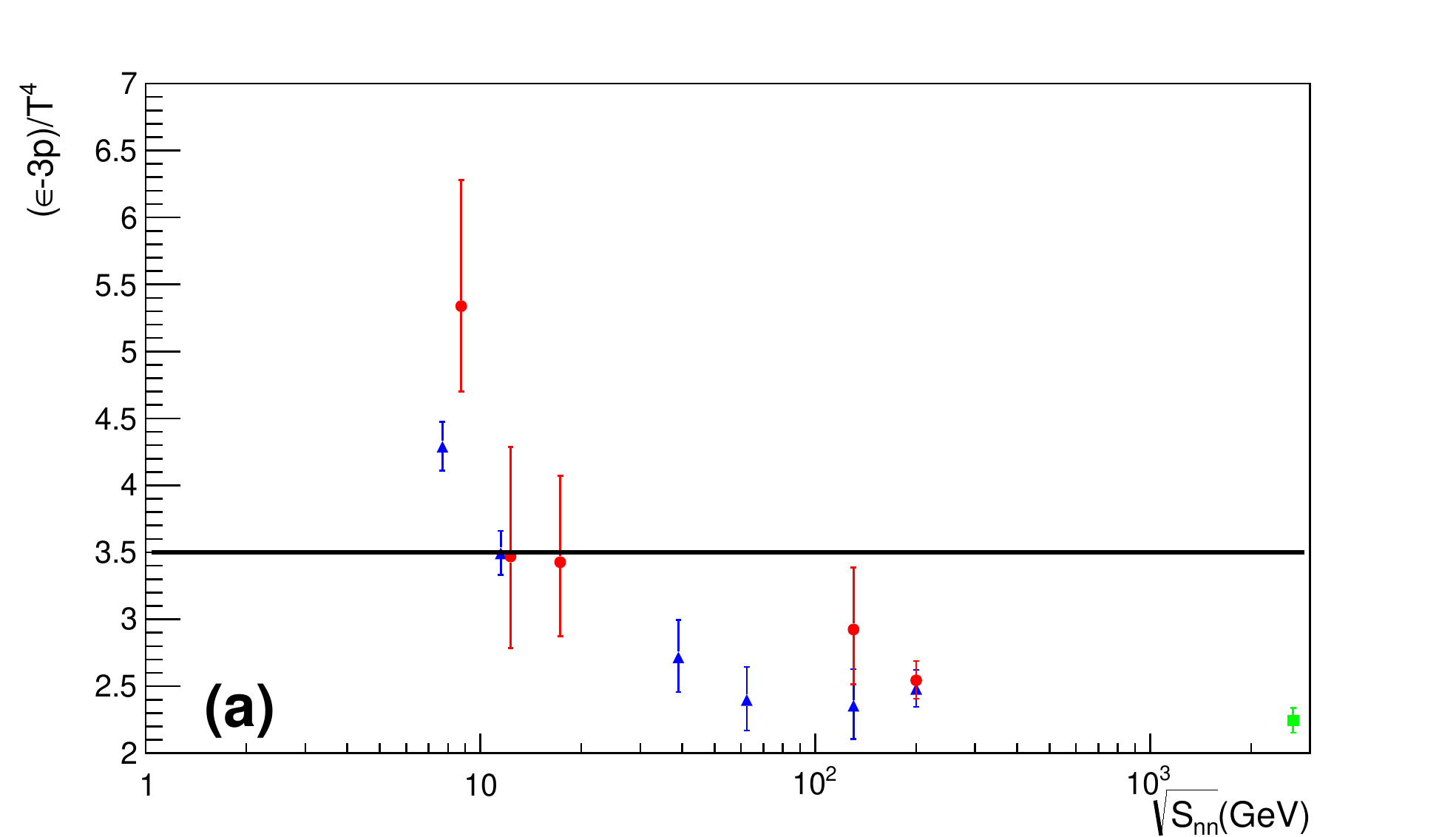}
\caption{The dependence of $s/n$ on energy calculated at the freezeout parameters $\mu_b$ and $T_{ch}$ ~\cite{Andronic2006,Tawfik2013,Stachel2014} is given. The closed triangle symbols are results from Ref.~\cite{Tawfik2013}, the circles stand for results from Ref.~\cite{Andronic2006} and the square is the results at LHC energy \cite{Stachel2014}. The solid line represents the constant value $(\epsilon-3p)/T^4=3.5$~\cite{Tawfik2013trace}. } 
\label{fig:trace_massL}  
\end{center}  
\end{figure}

Fig. \ref{fig:trace_massL} depicts $(\epsilon-3p)/T^4$ calculated from the extracted freezeout parameters at different energies by using GCE and full chemical equilibrium~\cite{Andronic2006,Tawfik2013,Stachel2014}. At low energy, it is obvious that $(\epsilon-3p)/T^4$ increases. At high energy ($>40~$GeV), this condition takes a constant value. Thus, the ratio $(\epsilon-3p)/T^4$ becomes no longer constant during chemical freeze-out at very large $\mu_b$. 
        
\begin{figure} 
\begin{center} 
\includegraphics[scale=0.4]{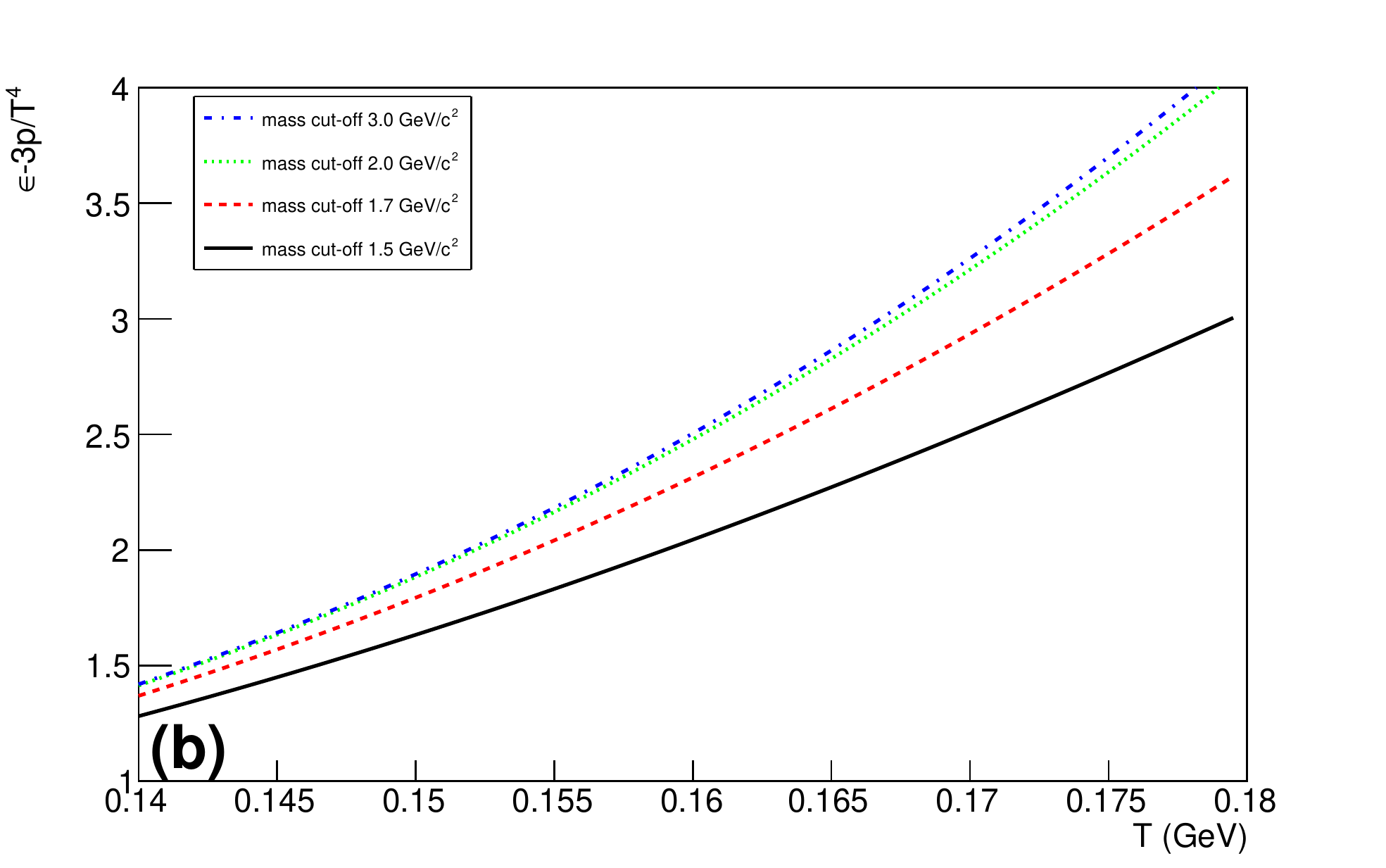}
\includegraphics[scale=0.43]{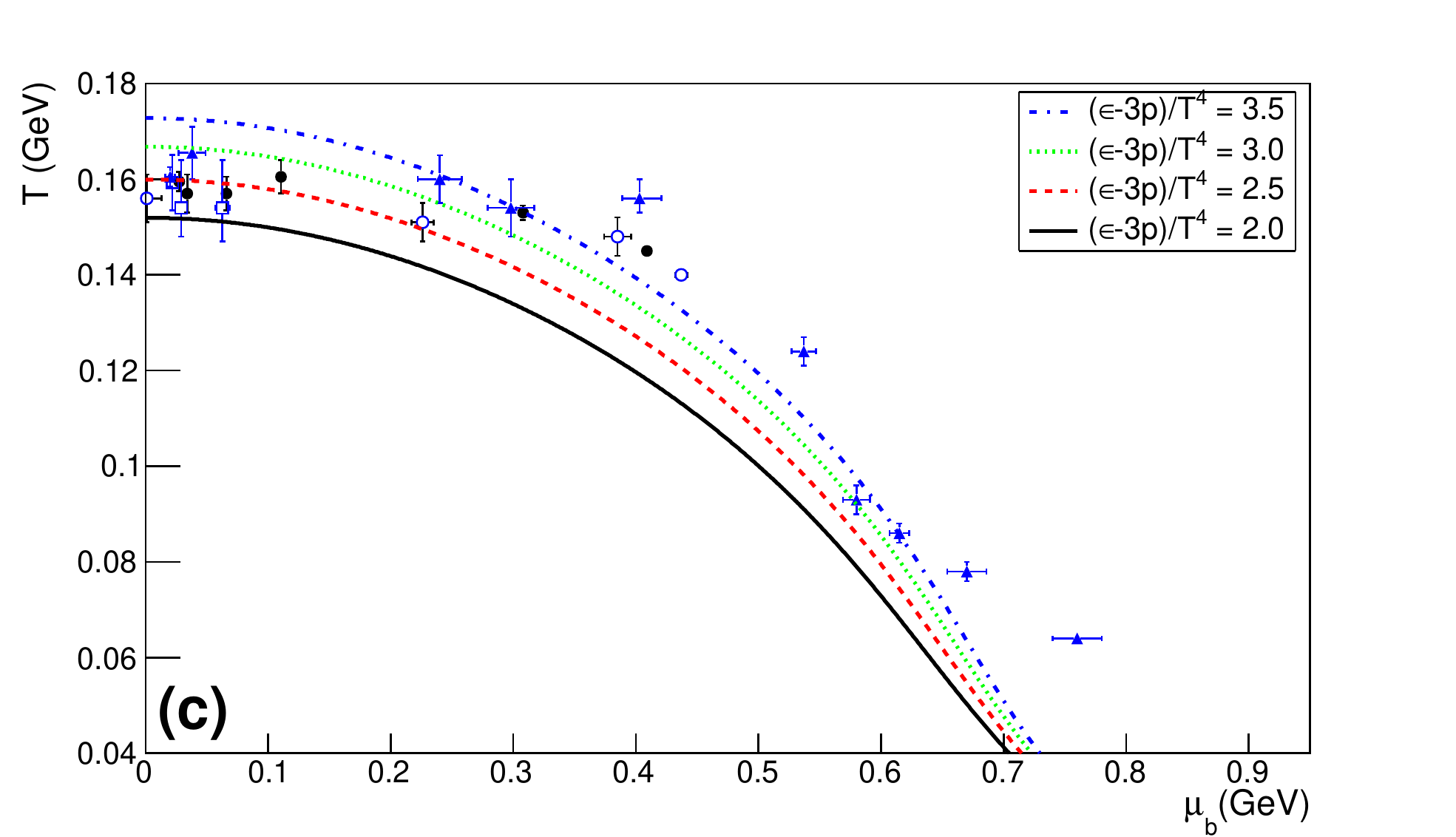}
\caption{Left-hand panel (b) shows the dependence of $(\epsilon-3p)/T^4$ on temperature at different resonance masses. Right-hand panel (c) gives freeze-out parameters $T_{ch}$ and $\mu_b$ determined as different $(\epsilon-3p)/T^4$ and compared with the ones deduced from particle ratios. The closed diamond symbols are results from Ref.~\cite{Andronic2006}, the open circles stand for results from Ref. \cite{Becattini2013}, the open squares represent the STAR results at $0-5\%$ centrality \cite{STAR2} and the closed circles are our results~\cite{Tawfik2013}. } 
\label{fig:trace_mass}  
\end{center}  
\end{figure}

In BM limit, the trace anomaly can be written as 
\begin{equation} 
\frac{\epsilon-3p}{T^4} = \sum_i\, g_i\, \left(\frac{m_i}{T}\right)^3\; K_1\, \left(\frac{m_i}{T}\right)\; \exp\left(\frac{\mu_i}{T}\right). \label{eq:CFOC5} 
\end{equation}               
In Eq. (\ref{eq:CFOC5}), when the mass cut is increased the terms that will be summed increase and the quantity $(\epsilon-3p)/T^4$ increases too. At different masses, different values of $T_{ch}$ are able to fulfil conditions of constant trance anomaly, left-hand panel of Fig. \ref{fig:trace_mass} (b). The EVC changes the value of $(\epsilon-3p)/T^4$ but not the idea that represents. In other words, when ECV is implemented one might find another guess for $(\epsilon-3p)/T^4$.

From recent lattice QCD calculations \cite{LQCDEoS}, $(\epsilon-3p)/T^4$ can be determined between $2.09$ and $2.76$. This is corresponding to  temperature ranging between $150$ and $160~$MeV. As the temperature decreases, higher values of $(\epsilon-3p)/T^4$ are needed. This is illustrated in right-hand panel of As Fig. \ref{fig:trace_mass} (c).

\section{Interrelations among chemical freeze-out conditions}
\label{sec:connect}

We found that at small baryon chemical potential, the conditions $s/T^3=7$ seems to coincide with $\epsilon/n=1.08~$GeV  \cite{cleymans2006}. Both conditions have been interpreted by Hawking-Unruh mechanism \cite{Castorina2007} for the particle production in high-energy collisions.  Almost identical values, $s/T^3=7.4$  and $\epsilon/n=1.09~$GeV have been determined \cite{Castorina2014,Tawfik:2015fda}. Furthermore, at high energies, the freeze-out conditions $s/n$ and $s/T^3$ both have almost the same value. This leads to a kind of interconnection between these different chemical  conditions.
\begin{enumerate}
\item It is obvious $\epsilon/n$ and $s/n$ can be related to each other 
\begin{eqnarray}
\sum_i \epsilon_i &=& T\, \sum_i s_i + \sum_i\, \mu_i\, n_i - \sum_i p_i, \\
\frac{\epsilon}{n} &=& T\, \left(\frac{s}{n} - 1\right) + \frac{\sum_i \mu_i n_i}{\sum_i n_i},
\label{eq:ensn} 
\end{eqnarray}
using $\sum_i p_i/\sum_i n_i= T$ in MB approximation.

\begin{figure} 
\begin{center} 
\includegraphics[scale=0.42]{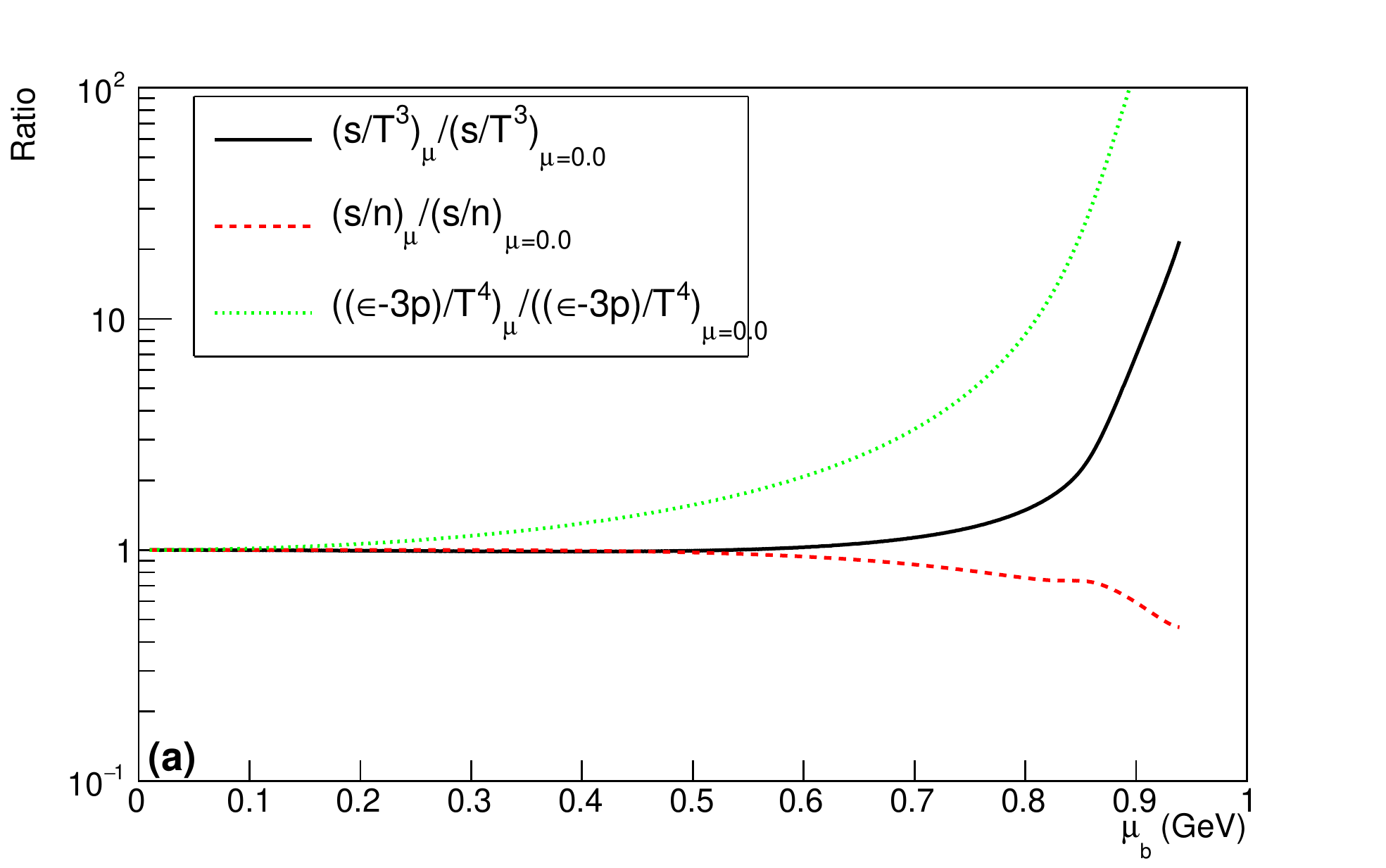}
\includegraphics[scale=0.42]{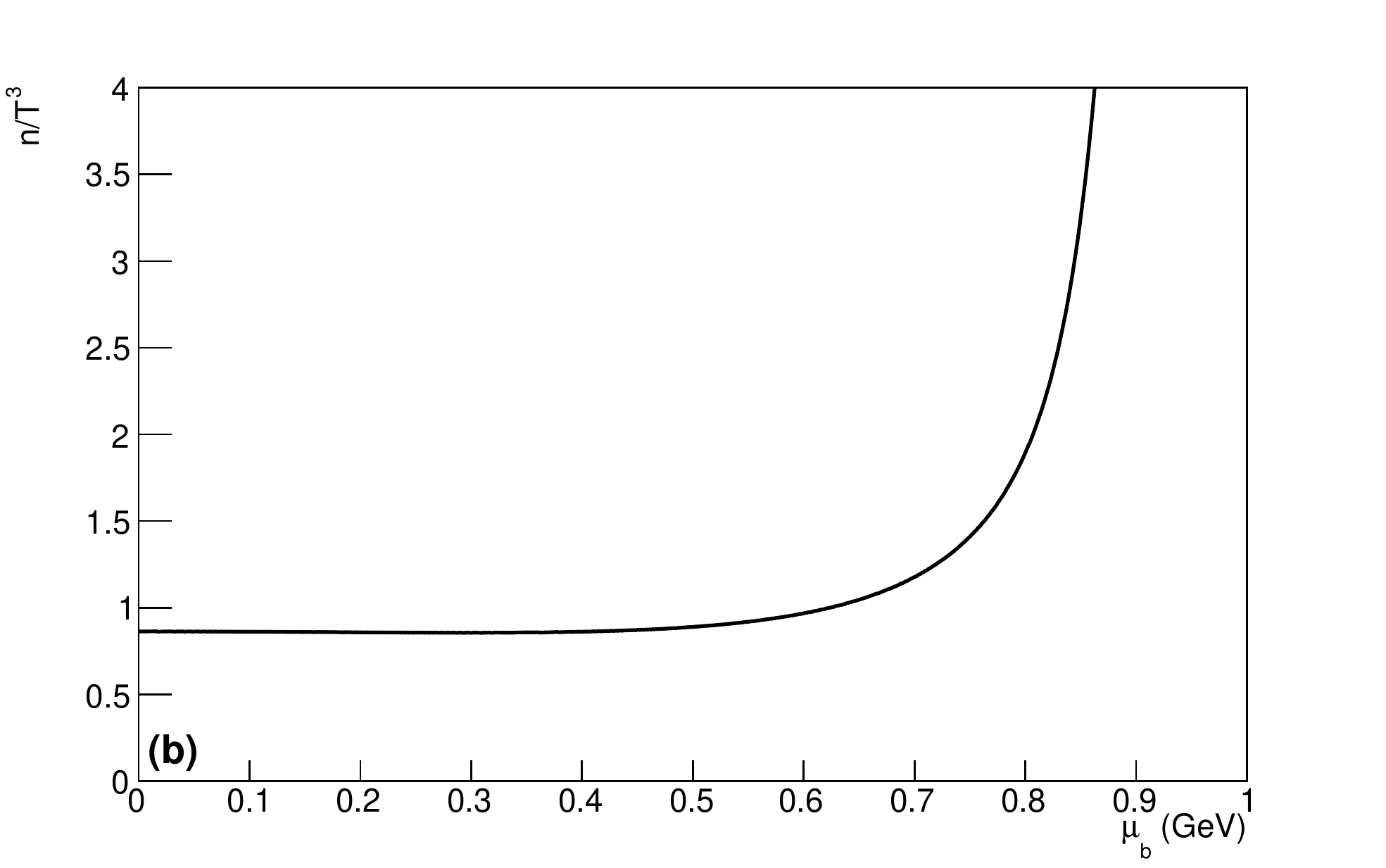}
\caption{\footnotesize Left-hand panel (a) shows the stability of $s/T^3$, $s/n$ and $(\epsilon-3p)/T^4$ normalized to the same quantity but at $\mu_b=0$ is given as a function of $\mu_b$. Right-hand panel (b) depicts the dependence of  $n/T^3$ on $\mu_b$. }
\label{fig:stability}  
\end{center}  
\end{figure} 

At small baryon chemical potential, let us assume $T=0.16~$GeV, for instance, then from the condition $\epsilon/n=1~$GeV, we get $s/n=7.25$. This means that both conditions are thermodynamically equivalent. But with increasing $\mu_b$, $s/n$ becomes no longer constant even at constant $\epsilon/n$. The dependence of $s/n$ at finite $\mu_b$ normalized to the corresponding value at vanishing $\mu_b$ at a fixed $\epsilon/n$ is depicted in left-hand panel of Fig. \ref{fig:stability} (a). When $\mu_b$ increases, $n/T^3$ does remain constant, [right-hand panel of Eq. \ref{eq:ensn} (b)]. Accordingly, the temperature varies. This leads to a departure in the value of the freezeout conditions with respect to their values at vanishing $\mu_b$ at constant $\epsilon/n$. In other words, both conditions can not be achieved, simultaneously. Although, there is a small deviation in $s/n$, which can be interpreted due to the huge change in $T_{ch}$ as predicted by $s/n$ condition at a certain $\mu_b$ [right-hand panel of Fig. \ref{fig:sn_mass} (c)].

\item  In MB limit, the pion number density is given as
\begin{equation}
n_{\pi} = \frac{g_{\pi}}{2 \pi^2} T m^2_{\pi} K_2\left(\frac{m_{\pi}}{T}\right)\; \exp\left(\frac{\mu_{\pi}}{T}\right).
\end{equation}
In non-relativistic limit, i.e. $m_{\pi}<<T$ and when ignoring the chemical potential and assuming that $K_2(m_{\pi}/T)$ can be approximated as $2 T^2/m_\pi^2$, then
\begin{equation}
n_{\pi} \approx \frac{3}{\pi^2} T^3.
\end{equation} 
    At high energy, the pions become dominating the formed fireball. Assuming that $n/T^3$ remains constant in high-energy collisions, then it is expected that $s/T^3$  turns to be related to $s/n$ as long as $n/T^3$ remains constant. The value of $n/T^3$ is approximated to $0.86$ at $\epsilon/n=1~$GeV as shown in right-hand panel of Fig. \ref{fig:stability} (b). At small baryon chemical potential, both conditions ($s/n$ and  $s/T^3$) are equivalent. At large baryon chemical potential, this equivalence seems to be destroyed. Thus, $s/n$, $s/T^3$ and $\epsilon/n$ become constants at small chemical potential.
    
\item In MB limit, $s/T^3$ and $(\epsilon-3p)/T^{4}$ can be related to each other,  
\begin{eqnarray}
\frac{s}{T^3} &=& \frac{1}{T^4} \sum_i \left(p_i + \epsilon_i - T\, \mu_i\, n_i \right), \\
\frac{\epsilon-3p}{T^{4}} &=& \frac{\sum_i s_i}{T^{3}} - \frac{4}{T^3} \sum_i n_i (1 - 0.25 \mu_i).
\end{eqnarray}
As discussed earlier, at small baryon chemical potential, $n/T^3$ is approximately constant. This leads to $\sum_i (\epsilon_i -3p_i)/T^4  \approx 3.56$.   
\end{enumerate} 

The summary of three interrelations is illustrated in Fig. \ref{fig:stability}. At small baryon chemical potential, $s/T^3$, $s/n$, $(\epsilon-3p)/T^{4}$ and $n/T^3$ are approximately constant at constant $\epsilon/n$. This conclusion does not contradict the results depicted in Figs. \ref{fig:en_massL}, \ref{fig:nbbar_massL}, \ref{fig:ST_massL}, \ref{fig:sn_massL} and \ref{fig:trace_massL}, as the constancy in  $s/T^3$, $s/n$, and $(\epsilon-3p)/T^{4}$ is based in their normalization to their corresponding value at $\mu=0$. But, as the baryon chemical potential increases, these freeze-out conditions become energy-dependent. In high-energy collisions, the thermodynamic quantities are weakly effected by the chemical potentials and thus remain approximately constant as well as the ratios between each pair of them.

\section{Properties of fireball thermodynamics at chemical freeze-out}
\label{sec:fireball}

It is assumed that the chemically equilibrated hadron gas emerges from the fireball produced in the high-energy collision. Determining its thermodynamic properties such as temperature, energy density, entropy etc. represents a great challenge to be related to the measurable properties in the final state, such as rapidity, momentum and the hadron multiplicity. As discussed in earlier sections, the hadron multiplicities and their ratios can be explained in terms of chemical freeze-out parameters, $T_{ch}$ and $\mu_b$, of the fireball. In the present work, it intends to redefine certain uniform conditions, e.g. chemical freeze-out conditions, which should be also satisfied by the thermodynamic properties of the fireball at the stage of the chemical freeze-out. In section \ref{sec:CFOC}, we have compared the extracted $T_{ch}$ and $\mu_b$ from the measured particle ratios with the parameters obtained from different freeze-out conditions. We try to highlight some details about the physical properties of the fireball. The ultimate goal is the characterization of the dynamics of the fireball expansion and how this can be accessed by the given chemical freeze-out conditions.

Here, we propose another way to check the chemical freeze-out conditions. We show that their values can be obtained from the extracted freeze-out parameters $T_{ch}$ and $\mu_b$~\cite{Tawfik2013}, which in this case represent the thermodynamic properties of the fireball at full chemical equilibrium and in most central collisions at mid-rapidity. From $T_{ch}$ and $\mu_b$ \cite{Tawfik2013}, we find for instance that the resulting temperature agrees well with the lattice QCD pseudo-critical temperature. 

Furthermore, from  the HRG model at $\mu_b \le 0.4~$GeV, following thermodynamic quantities can be determined
\begin{itemize} 
\item pressure ($p$) lies between $0.057$ and $0.065~$GeV/fm$^{-3}$,
\item energy density ($\epsilon$) ranges between  $0.393$ and $0.4312~$GeV/fm$^{-3}$, and
\item entropy density ($s$) extends between $2.68$ and $3.08~$fm$^{-3}$,
\end{itemize}
and accordingly the freeze-out conditions lead to
\begin{itemize}
\item energy per particle ($\epsilon/n$) lies between $0.95$ and $1.05~$GeV, 
\item entropy per particle ($s/n$) ranges between $6.94$ and $7.07$,
\item normalized entropy density ($s/T^3$) extends between $5.4$ and $6.8$, and 
\item trace anomaly [$(\epsilon-3p)/T^{4}$] takes a value within $2.5$ and $4$.
\end{itemize}

It is obvious that these values which are estimated for the chemical freeze-out conditions agree well with the values which are assigned to them, phenomenologically. Furthermore, small changes in the thermodynamic properties are noticed at $\mu_b \le 0.4~$GeV. In this $\mu_b$-range, the temperature does not vary so much. Additionally, we conclude that the real test of the chemical freeze-out conditions at low energy (large $\mu_b$) is the one where $T_{ch}$ and $\mu_b$ are changing very rapidly with changing $\sqrt{s_{NN}}$.  In this energy  (or $\mu_b$) range, the differentiation between the various chemical freeze-out conditions likely becomes obvious.

%%%%%%%%%%%%%%%%%%  Section VI %%%%%%%%%%%%%%%%%%
\section{Conclusions and Outlook}
\label{sec:con}

We have reanalysed some chemical freeze-out conditions in framework of HRG model and compared them with new ones. The calculations are performed in full chemical equilibrium in grand canonical ensemble with zero-width approximation. The excluded-volume corrections are not applied.
 
Most of conditions match well with the chemical freeze-out parameters, which are extracted from fitting the measured particle ratios with their calculations in the HRG model. In doing this, at least two free parameters, $T_{ch}$ and $\mu_b$ are to be tuned. This assures the conservation laws of strangeness, charge and baryon numbers and are to be controlled by the phase space. 

Some of these conditions are ratios between two extensive thermodynamic quantities, such as $s/n$ and $\epsilon/n$. Thus, it is almost identical to study them with or without excluded-volume corrections. But, others are effected by the excluded-volume corrections, such as $s/T^3$, $(\epsilon-3p)/T^{4}$ and $n_{b}+n_{\bar b}$. Here we assume that the radii of mesons and baryons are vanishing. The scope of this work is defining interrelations among various chemical freeze-out conditions. EVC probably comes up with corrections to various thermodynamic quantities. These likely should not change their interrelations. 
 
It is obvious that all conditions are effected by the mass cut-off, especially at small $\mu_b$, as given in the left-had panels of the first five figures. With increasing the resonance masses, their contributions increase and consequently lower temperatures are able to fulfil the given freeze-out conditions. Thus, at different resonance mass limits, different $T_{ch}$ are able to fulfil the condition.

Regarding the interrelations among the various chemical freeze-out conditions, we find that 
\begin{enumerate}
\item the condition $s/T^3=7$ coincides with $\epsilon/n=1.08~GeV$ \cite{cleymans2006}, especially at small baryon chemical potential. 
\item At small $\mu_b$, the two conditions $s/n$ and $s/T^3$  have approximately the same dimensionless value, i.e. $7$. It can be assume that these interrelations can be generalized to the whole range of chemical potentials. 
\item This would make it possible to consider the possibility to interpret the different conditions as different aspects of one universal condition.  At small or even vanishing $\mu_b$, $\epsilon/n=1~$GeV. This value leads to $s/n=7.25$. 
\item $s/T^3$ turns to be related to $s/n$ as long as $n/T^3$ remains constant.  
\item At small baryon chemical potential, a clear relation is found between trace anomaly, normalized entropy and $n/T^3$. As long as $n/T^3$ remains constant and $\mu_b$ is very small, $(\epsilon-3p)/T^{4} \approx 3.56$ and $s/T^3=7$ are found equivalent. 
\end{enumerate}

Most of freeze-out conditions coincide with the experimental measurements at high energy. This agreement becomes weaker, at lower energies. We would like to highlight that these conclusions should be restricted to the constrains made in this study; full chemical equilibrium, zero-width approximation and grand-canonical ensemble framework. But these collusions are correct at high energy. At lower energies, the freeze-out conditions are no longer dominant or precise in order to draw edge-cutting conclusion. The proposed interrelations among the various chemical freeze-out conditions, which have been so far achieved partly, lead to interpreting the different conditions as different aspects from fewer universal conditions..

Recent studies suggest that the chemical freeze-out occurs in a slightly non-equilibrium situation, for instance, due to hadron inelastic rescattering and non-equilibrium quark occupation factors ($\gamma_q$ and $\gamma_s$).  The latter is widely criticized due to the additional free parameters included in the statistical fits \cite{michelle}. The early one was also critically commented because of the absence of stringent test, especially with yields of light nuclei, in which no additional free parameter is available \cite{Stachel2014}. The light nuclei are sensitive to the increasing in the powers of the quark chemical potentials. For sake of completeness, we highlight that the production of hypertritons, for instance, is in good agreement with the standard statistical hadronization picture. But they are largely overpredicted (by a factor of $6$), at non-equilibrium $\gamma_q$ and $\gamma_s$.

%-----------------------------------------------

%----------------------==================-------------------------

\end{document}